\documentclass[amsmath,eqsecnum,preprint,pre,aps,showpacs]{revtex4}
\usepackage{amsmath}
\usepackage{graphicx}
\usepackage{dcolumn}
\usepackage{bm}
\usepackage{float}
\usepackage{xcolor}

\begin{document}

\draft
\emph{}
\title{Non-equilibrium theory of the linear viscoelasticity \\ of  glass and gel forming liquids.}
\author{R. Peredo-Ortiz$^{1,2}$, O. Joaqu\'in-Jaime$^{1}$,  L. L\'opez-Flores$^{3}$, M. Medina-Noyola$^{1}$ and L.F. Elizondo-Aguilera$^{4}$.}
\affiliation{$^{1}$ Instituto de F\'{\i}sica,
Universidad Aut\'{o}noma de San Luis Potos\'{\i}, \'{A}lvaro
Obreg\'{o}n 64, 78000 San Luis Potos\'{\i}, SLP, M\'{e}xico}
\affiliation{$^{2}$ Facultad de Ciencias F\'isico-Matem\'aticas,
Benem\'erita Universidad Aut\'{o}noma de Puebla, Apartado Postal 
1152, CP 72570, Puebla, PUE, M\'{e}xico}
\affiliation{$^{3}$ Department of Materials Science and Engineering, Northwestern University, 
Evanston, Illinois 60208, USA.}
\affiliation{$^4$ Instituto de F\'isica, Benem\'erita Universidad Aut\'onoma de Puebla, 
Apartado Postal J-48, 72570 Puebla, Mexico.}

\date{\today}

\begin{abstract}

We propose a first-principles theoretical approach for the description of the aging of the linear viscoelastic properties of a colloidal liquid after a sudden quench into a dynamically arrested (glass or gel) state. Specifically, we couple a general expression for the time-evolving shear-stress relaxation function $\eta(\tau;t)$ (whose $\tau$-integral is the instantaneous viscosity $\eta(t)$), written in terms of the non-equilibrium structure factor $S(k;t)$ and intermediate scattering function $F(k,\tau;t)$, with the equations that determine $S(k;t)$ and $F(k,\tau;t)$, provided by the non-equilibrium self-consistent generalized Langevin equation (NE-SCGLE) theory. In this manner, we obtain a closed theoretical scheme that directly connects inter-particle forces with experimentally accessible rheological properties of non-equilibrium amorphous states of matter. The predictive capability of the resulting theoretical formalism is illustrated here with its concrete application to the Weeks-Chandler-Andersen (WCA) model of a soft-sphere fluid.

\end{abstract}


\maketitle

\section{Introduction}\label{section1}

The physical understanding of the structural, dynamical and rheological behavior of gelled and glassy states is essential for the rational design of novel materials, and also, an important element in manufacturing many ordinary industrial products, such as fiber optic glasses, electronic devices, cosmetics, food, pharmaceuticals and metallic alloys \cite{mewiswagnerbook21,ChenJanmeyYodh}. The properties of gels and glasses, however, are strongly influenced by a complex interplay between dynamical arrest mechanisms, phase separation and aging phenomena, and hence, depend on both the observation time and the protocol of fabrication \cite{johari}, in clear contrast with ordinary equilibrium phases. In consequence, a comprehensive first-principles  description of the properties of these non-equilibrium amorphous states of matter is still missing, remaining as one of the major unsolved challenges for theoretical physics and condensed soft matter.

A crucial aspect of the rheology of gels and glasses is concerned, in particular, with the time-evolving relaxation of stresses during deformation, which causes gradual changes in mechanical properties such as viscosity, elasticity or plasticity, thus modifying their flow behavior \cite{voigtmann1,winkler}. The experimental determination of the kinetics of these processes, requires monitoring the dependence of specific measurable properties on the evolution time. For example, employing conventional rheometric techniques that involve the application of a small oscillatory shear strain of frequency $\omega$, it is possible to test the linear viscoelastic response of glass and gel-forming materials \cite{ferry}. Such response is encoded in the so-called storage and loss moduli, $G'(\omega;t)$ and $G''(\omega;t)$, which determine the macroscopic stress in the material induced by the strain at observation time $t$ \cite{purnomo,purnomo1,suman,christopoulou,mckenna1,mckenna2,mckenna3}. To better understand and control the rheological behavior of non-equilibrium gels and glasses, however, one also requires a first-principles description of its microscopic origin, ideally based on theories capable of explaining viscoelastic response in terms of inter-particle forces, macroscopic control parameters and fabrication protocols.  


In this context, a starting point for the theoretical description of the linear viscoelasticity of glass-forming
atomic liquids was provided by Geszti \cite{geszti}. This initial work was extended to colloidal suspension by Naegele and Bergenholtz \cite{naegele,banchio}, who built a rigorous self-consistent mode-coupling scheme that relates the linear viscoelastic properties of a dense colloidal liquid with diffusion mechanisms. Such scheme, however, leads to a diverging shear viscosity at the glass transition (GT) point, as a reminiscence of the divergence of the structural relaxation time $\tau_{\alpha}$ predicted by the mode coupling theory (MCT) \cite{goetze1}, an idealized physical scenario that is never observed in practice. Hence, notwithstanding this approach provides a good reference framework to investigate many important signatures of the viscoelastic behavior of a liquid in the approach to the GT, it should be noted that it is intrinsically restricted (just as MCT) to the description of systems that are ``nearly-arrested", but still in thermodynamic equilibrium. In consequence, the study of the fundamental kinetic fingerprints of dynamically arrested states, such as the aging of the viscoelastic moduli $G'(\omega;t)$ and $G''(\omega;t)$, observed in careful and systematic experimental measurements  \cite{purnomo,suman,christopoulou,mckenna1,mckenna2,mckenna3}, is definitely out of the reach of the above theoretical proposal. It is thus important to extend the available theoretical tools \cite{naegele,banchio,voigtmann1,brader,brady} to understand non-equilibrium aging phenomena, at the level of these rheological properties.

The present work is aimed at addressing this theoretical challenge. More specifically, here we present a first-principles theory, i.e., a theory rooted in the particle-particle interaction forces, able to describe the aging processes that occur in the linear viscoelastic properties of a simple glass- or gel-forming liquid, after a sudden quench into its regions of dynamical arrest. The specific predictions and the resulting scenario that emerge from this novel theory will be illustrated in detail with its application to a simple soft-sphere model fluid. Here we shall be particularly interested in describing the aging of such a liquid by monitoring the $t$-evolution of the total shear-stress relaxation function $\eta(\tau;t)$, whose  integral $\eta(t)\equiv \int_0^\infty \eta(\tau;t) d \tau$ is the instantaneous value of the ordinary viscosity. The Fourier-Laplace transform $\eta(\omega;t)\equiv  \int_0^\infty e^{-i\omega \tau}\eta(\tau;t) d\tau$  is the dynamic shear viscosity \cite{christopoulou,mewiswagnerbook21}, related with the complex-valued dynamic shear modulus $G(\omega;t)$ by the relationship $G(\omega;t) =  i\omega \eta(\omega;t)$, where the real and imaginary parts of $G(\omega;t)$ are precisely the aforementioned elastic and loss moduli, $G'(\omega;t)$ and $G''(\omega;t)$. 
The approximate first-principles theoretical protocol proposed here to determine these $\omega$-dependent \emph{linear} viscoelastic properties, at any finite waiting time $t$, will have in mind a generic glass- or gel-forming liquid, i.e., a monocomponent systems of particles interacting through any radially-symmetric inter-particle pair potential $u(r)$, instantaneously quenched from arbitrary initial conditions, to arbitrary final temperature and/or density. 

Our strategy to construct this first-principles theoretical protocol will be to mimic its equilibrium counterpart \cite{naegele}, which connects the pair potential $u(r)$ with the rheological properties in three distinct steps. The first connects $u(r)$ with the equilibrium static structure factor $S(k)$ by means of well-established approximate theories of liquids \cite{mcquarrie,hansen}. The second connects $S(k)$ with  equilibrium dynamic properties, namely, the collective and self intermediate scattering functions $F(k,\tau)$ and  $F_S(k,\tau)$, as, for example, in MCT \cite{goetze1}. Finally, the third expresses the viscoelastic properties in terms of  $S(k)$ and  $F(k,\tau)$, as done by Geszti \cite{geszti} and by Naegele and Bergenholtz \cite{naegele,banchio} (see, e.g., Eq. (40) of Ref. \cite{banchio}). 

In principle, to extend this strategy to non-equilibrium conditions, we require the extended version of  each of these three steps. As we explain in detail in Section \ref{section2}, when extended to non equilibrium, the first two steps (which connect  the pair potential $u(r)$ with the structure, and then with the dynamics) are integrated in a single grand connection, provided by  the recently-developed statistical physical formalism referred to as the  \emph{non-equilibrium self-consistent generalized Langevin equation} (NE-SCGLE) theory \cite{nescgle1,nescgle2,nescgle3}. The essence of this theory is a set of time-evolution equations for the non-equilibrium structure factor $S(k;t)$ and the collective and self  intermediate scattering functions (ISFs)  $F(k,\tau;t)$ and $F_S(k,\tau;t)$ (Eqs. \eqref{relsigmadif2appendix}-\eqref{lambdadk} below), given the pair potential $u(r)$ and the preparation protocol \cite{gabriel,nescgle5,mendoza}.

In contrast, to our knowledge, there is no extension to non-equilibrium conditions, of the third step, i.e., no  expression is known for the non-equilibrium viscoelastic properties in terms of  $S(k;t)$ and $F(k,\tau;t)$, analogous to the equilibrium expression derived by Geszti \cite{geszti} and by Naegele and Bergenholtz \cite{naegele,banchio}. Thus, the first goal of the present work is to establish this missing extension, such that, combined with the NE-SCGLE equations for $S(k;t)$, $F(k,\tau;t)$, and $F_S(k,\tau;t)$, becomes a general protocol for the theoretical prediction from first principles (i.e., for given $u(r)$) of the aging of the viscoelastic properties of a non-equilibrium liquid. Then, the second goal of this work is to   illustrate the use of this protocol by means of a concrete application. For this, we use it to describe the aging of the viscoelastic response  of a generic model of a soft-sphere fluid, the Weeks-Chandler-Andersen (WCA) model, after being instantaneously quenched into a glassy state.

Our non-equilibrium expression for the viscoelastic properties in terms of the structure and dynamics of the system is written in Eq. (\ref{deltaeta}) below. One immediately notices its similarity with the expression derived by  N\"agele and Bergenholtz \cite{naegele,banchio} for Brownian liquids at thermodynamic equilibrium conditions. This coincidence, however, is not accidental, since in our derivation of Eq. (\ref{deltaeta}) we adopted a similar line of arguments, simplifications, and approximations. We only checked  that these arguments and approximations  in reality never required the condition of thermodynamic equilibrium, but only the mathematical condition of stationarity or other temporal or spatial symmetries, or even some general mathematical approximations, such as the Gaussian factorization of four-point correlations.

This strategy, in fact, seems to be a useful procedure to extend to non-equilibrium, some of the well-established  theoretical relations and results of the equilibrium statistical mechanics of classical fluids, such as the energy equation  \cite{mcquarrie,hansen} or the Wertheim-Lovett relation  \cite{evans}. This is, of course, a relevant methodological subject of more general interest in the study of liquids under non-equilibrium conditions.   For this reason,  we address this subject in a separate manuscript \cite{wertheimlovett}, in which we include the detailed derivation of  Eq. (\ref{deltaeta}) as an illustrative example of such general methodology. This allows us to concentrate the present work mostly on the use of this expression, together with the NE-SCGLE equations, as a theoretical protocol for the description of the non-equilibrium linear viscoelasticity. 

This work is organized as follows. In Section \ref{section2} we briefly outline the derivation of our non-equilibrium extension of the Geszti/Naegele-Bergenholtz expression for the time-evolving shear-stress relaxation function $\Delta\eta(\tau;t)$ in terms of $S(k;t)$ and $F(k,\tau;t)$ (where $t$ is the ``waiting-time'' after the quench). We then briefly review the NE-SCGLE theory, whose essence are the time-evolution equations for $S(k;t)$, $F(k,\tau;t)$,  and $F_S(k,\tau;t)$. Combined, these results constitute the general protocol for the description of the non-equilibrium evolution (aging) of the viscoelastic properties of a non-equilibrium liquid.

The availability of this general protocol opens the door to a wide variety of possibilities to model virtually any aspect of the non-equilibrium linear viscoelasticity of ``simple'' glass-forming liquids. Thus, after establishing this general protocol in Section \ref{section2}, the rest of this work is devoted to describe some of these possibilities in the context of its application to the simple Weeks-Chandler-Andersen (WCA) model \cite{wca} of a soft-sphere fluid. For this, in Section \ref{section3} we describe the  non-equilibrium evolution of  the shear-stress relaxation function $\Delta\eta(\tau;t)$ and of the total shear viscosity $\eta(t)$ for an individual quench, comparing the process of equilibration with the aging process of dynamic arrest. In contrast, in Section \ref{section4} we describe the collective evolution of an ensemble of quench processes with a distribution of final densities and temperatures. In the simplest distribution studied we fix the volume fraction, thus generating  the essence of a $t$-dependent  version of the well-known Angell plots \cite{angell}. We then consider a more general exercise in which the volume fraction is no longer fixed, to discuss the recently-developed concept of time-dependent glass-transition diagrams \cite{zepeda} in the simpler context of the WCA model. Finally, Section \ref{section5} summarizes the main findings of this work and its perspectives.

\section{Non-equilibrium linear viscoelasticity of a colloidal liquid}\label{section2}

Let us now discuss the main aim of this work which, as just mentioned, is to propose a theoretical framework for the first-principles description of the non-equilibrium linear viscoelasticity of a fluid after a sudden quench into a glass state. For clarity, it is instructive to start by reviewing some pertinent definitions and general relations  involving the linear viscoelastic response of a colloidal liquid in the absence of hydrodynamic interactions \cite{naegele,banchio}.

\subsection{Linear Viscoelasticity: General relations}

Let us consider a colloidal suspension of $N$ identical spherical particles with diameter $\sigma$ in a volume $V$, interacting through a radially-symmetric pairwise potential $u(r)$ and subjected to the action of a weak oscillatory shear flow of frequency $\omega$ and shear rate amplitude $\dot{\gamma}_0$ \cite{mewiswagnerbook21,ferry}. The fluid flow velocity is assumed to be given by the real part of  $\mathbf{u}(\mathbf{r},\tau)=\dot{\gamma}_0y\hat{\mathbf{x}}e^{i\omega \tau}$, where $\hat{\mathbf{x}}$ is the unit vector in the $x$-direction. For simplicity, we shall only consider the limit of sufficiently small shear rate amplitudes, $\dot{\gamma}_0 \to 0$, in which the isotropic and homogeneous structure of the suspension is not significantly distorted. In this limit, the linear relationship ${\bm{\Sigma}}(t)=\int_{0}^{t}dt'\eta(t,t')\mathbf{E}(t')$ between the macroscopic stress tensor ${\bm{\Sigma}}(t)$ and the rate of strain tensor $\mathbf{E}(t)$ constitutes the phenomenological definition of the  isotropic and homogeneous total shear stress relaxation function $\eta(t,t')$. Under stationary conditions, ${\bm{\Sigma}}(t)$ and $\mathbf{E}(t)$ become the constants ${\bm{\Sigma}}$ and $\mathbf{E}$, and $\eta(t,t')$ becomes $\eta(t,t')=\eta(t-t')$, so that the linear relationship above becomes ${\bm{\Sigma}}= \eta \mathbf{E}$, with the constant  $\eta$ being the ordinary macroscopic viscosity $\eta\equiv \int_0^\infty \eta(\tau)d\tau$.

In what follows, however, we shall consider our homogeneous suspension to be initially in thermal equilibrium during the time  $-\infty < t \le 0$ at an initial density $n_i=N/V$ and temperature $T_i$. This system is then  subjected at $t=0$ to an instantaneous quench in control parameters to the final values $n$ and $T$. In response, the system  must adjust itself over the time $t>0$ to new stationary conditions. During this relaxation transient, the non-stationary total shear stress relaxation function $\eta(t,t')$ can be written as $\eta(t,t')=\eta(t-t';t)$, i.e., as $\eta(\tau;t)$, with $\tau\equiv t-t'$. This, in fact, is the non-equilibrium  shear stress relaxation function referred to in the introduction, whose Fourier-Laplace transform $\eta(\omega;t)$  is the dynamic shear viscosity, related with the dynamic shear modulus $G(\omega;t)$ by  $G(\omega;t) =  i\omega \eta(\omega;t)$, whose real and imaginary parts are the elastic and loss moduli $G'(\omega;t)$ and $G''(\omega;t)$.


The total shear stress relaxation function, $\eta(\tau;t)$, can be written as $\eta (\tau;t)= 2\delta (\tau) \eta^0 + \Delta \eta (\tau;t)$, with $\eta^0$ being the ``short-time'' (or ``infinite-frequency'') viscosity and $D^0$ the ``short-time'' (or``free'') self-diffusion coefficient, related with each other by the Stokes-Einstein relation $\eta^0= k_BT/3\pi\sigma D^0$. The function $\Delta\eta (\tau;t)$, in turn, is the contribution to $\eta (\tau;t)$ due to the inter-particle forces. In the absence of hydrodynamic interactions, $\eta^0$ is the viscosity of the pure solvent, but under some circumstances, such as for concentrated hard-sphere suspensions, the effects of hydrodynamic interactions act virtually instantaneously, simply renormalizing the value of $\eta^0$ and $D^0$, but otherwise behaving as if hydrodynamic interactions were absent \cite{prlhi,mazurgeigen}. This will be a general assumption in what follows.

\subsection{Expression for $\Delta \eta (\tau;t)$.}


We now describe the derivation of our non-equilibrium expression for the shear-stress relaxation function $\Delta\eta(\tau;t)$ in terms of the time-dependent structural and dynamical properties $S(k;t)$ and $F(k,\tau;t)$. Our starting point is the following equation, 
\begin{equation}
\Delta \eta (\tau;t)= (\beta/V)\langle \sigma ^{xy}(t+\tau) \sigma ^{xy}(t)\rangle, 
\label{greenkubo0}
\end{equation}
in which $\sigma ^{xy}(t)$ denotes the microscopic expression for the configurational component of  the stress tensor, given by 
\begin{equation}\label{eq: sigma1}
\sigma^{xy}(t)  = -\hat{\mathbf{x}}\cdot \left[ \sum_{i=1}^N\mathbf{r}_i(t)\mathbf{F}_i(t)\right] \cdot\hat{\mathbf{y}},
\end{equation}
where $\mathbf{r}_i(t)$ and $\mathbf{F}_i(t)$ are the position and total force on the $i$-th colloidal particle, respectively.

Eq. (\ref{greenkubo0}) is a well-known example of a Green-Kubo relation, normally proposed and applied assuming thermodynamic equilibrium conditions \cite{hansen,naegele,boonyip}, in which case  the brackets $\langle ... \rangle$ indicate an ordinary \emph{equilibrium} (canonical, microcanonical, ...) ensemble average and $\beta\equiv 1/k_BT$, where $T$ is the temperature, which must be the same at any position within the system, and equal to the temperature of the reservoir, $T=T^R$. In what follows, however, we shall not assume thermodynamic equilibrium, and hence, the brackets $\langle ... \rangle$ will indicate average over an arbitrary  \emph{non-equilibrium} (and even non-stationary) ensemble. In principle, this means that $(k_B\beta)^{-1}$ is no longer a constant temperature $T$. Instead, in general $\beta^{-1}$ is defined as  $\beta^{-1}=k_BT(\mathbf{r};t)\equiv   \langle \mathbf{p}^2(t)/3M\rangle$, where  $\mathbf{p}(t)$ is the momentum of one representative particle, whose mean kinetic energy defines the  space- and  time-dependent molecular temperature $T(\mathbf{r};t)$ \cite{wertheimlovett}. For simplicity, however, we shall neglect the space dependence of the molecular temperature, $T(\mathbf{r};t)=T(t)$, which implies an instantaneous thermalization of the degrees of freedom associated with the kinetic energy of the particles (i.e., infinite heat conductivity). Furthermore, we shall assume that $T(t)$ will  instantaneously assume the value $T^R(t)$ imposed by the reservoir. Thus, the temperature becomes an external control parameter, with the externally imposed time-dependence of $T(t)=T^R(t)$ representing the thermal preparation protocol. From now, we shall only consider the simplest protocol, namely, an instantaneous temperature quench from an initial temperature $T_i$ to a final temperature $T_f$, occurring at time $t=0$. With these considerations in mind, for $t>0$, we can write $\beta\equiv 1/k_BT_f$. For notational convenience, however, in what follows $T_f$ will be denoted simply by $T$. 

In the absence of external fields, we may rewrite equation \eqref{eq: sigma1} as
\begin{equation}
 \sigma ^{xy}(t) = -\sum_{i=1}^N R_i^x(t) F_i^y(t) = \frac{1}{2}\sum_{ i, j =1}^N x_{ij}\frac{du(R_{ij})}{dy_{ij}},
\label{defsigmaxy}
\end{equation}
where $R_i^x(t) \equiv \hat{\mathbf{x}}\cdot \mathbf{r}_i(t)=x_i(t)$, $F_i^y(t)\equiv \mathbf{F}_i(t) \cdot\hat{\mathbf{y}}$, $x_{ij}(t)=x_i(t)-x_j(t)$, and ${du(R_{ij})/dy_{ij}}\equiv (\nabla_{ij} u(R_{ij})) \cdot \hat{\mathbf{y}}$. In terms of the local density of particles 
\begin{equation}
n(\mathbf{r},t)\equiv \sum_{i=1}^N \delta(\mathbf{r}-\mathbf{r}_i(t)), 
\label{defndrt}
\end{equation}
one can rewrite Eq. \eqref{defsigmaxy} as 
\begin{eqnarray}
 \sigma ^{xy}(t)&=&\frac{1}{2}\int d\mathbf{r} \int d\mathbf{r}' (x-x') \frac{du(\mid \mathbf{r}-\mathbf{r}'\mid)}{d(y-y')} \sum_{i=1}^N \delta(\mathbf{r}-\mathbf{r}_i(t))\sum_{ j=1}^N \delta(\mathbf{r}'-\mathbf{r}_j(t))\nonumber \\
&=&\frac{1}{2}\int d\mathbf{r} \int d\mathbf{r}' (x-x') \frac{du(\mid \mathbf{r}-\mathbf{r}'\mid)}{d(y-y')} n(\mathbf{r},t)n(\mathbf{r}',t), 
\label{defsigmaxy1} 
\end{eqnarray}
which, inserted  in Eq. (\ref{greenkubo0}), leads to
\begin{eqnarray}
\Delta \eta (\tau;t)&=& \frac{(\beta/V) }{4}\int d\mathbf{r}_1  d\mathbf{r}_2 d\mathbf{r}_3 d\mathbf{r}_4(x_1-x_2) \frac{du(\mid \mathbf{r}_1-\mathbf{r}_2\mid)}{d(y_1-y_2)} (x_3-x_4) \frac{du(\mid \mathbf{r}_3-\mathbf{r}_4\mid)}{d(y_3-y_4)}  \nonumber \\
& & \times \Big\langle n(\mathbf{r}_1,t+\tau)n(\mathbf{r}_2,t+\tau)n(\mathbf{r}_3,t)n(\mathbf{r}_4,t)\Big\rangle.
 \label{deltaeta1}
\end{eqnarray}

Using  the Fourier transform $u(k) = \int d\mathbf{r} e^{-i\mathbf{k}\cdot \mathbf{r}} u(r)$ of the pair potential $u(r)$, it is straightforward to show that $-\mathbf{r}\nabla u(r) = (1/(2\pi)^3)\int d\mathbf{k} e^{i\mathbf{k}\cdot \mathbf{r}} \nabla_\mathbf{k}[\mathbf{k}u(k)]$, whose $xy$ component (writing $\mathbf{r}=\mathbf{r}_1-\mathbf{r}_2$) is
\begin{equation}
(x_1-x_2) \frac{du(\mid \mathbf{r}_1-\mathbf{r}_2\mid)}{d(y_1-y_2)} = \frac{-1}{(2\pi)^3}\int d\mathbf{k} e^{i\mathbf{k}\cdot ( \mathbf{r}_1-\mathbf{r}_2)} \hat{\mathbf{x}} \cdot \nabla_\mathbf{k}[\mathbf{k} \cdot \hat{\mathbf{y}} u(k)] = \frac{-1}{(2\pi)^3}\int d\mathbf{k} e^{i\mathbf{k}\cdot ( \mathbf{r}_1-\mathbf{r}_2)} \left( \frac{\partial \left[k_yu(k)\right]}{\partial k_x} \right).
\label{greenkubo}
\end{equation}
This result allows us to rewrite Eq. (\ref{deltaeta1}) as 
\begin{eqnarray}
\Delta \eta (\tau;t)&=& \frac{(\beta/V) }{4(2\pi)^6}\int d\mathbf{k}\int d\mathbf{k}' \left( \frac{\partial \left[k_yu(k)\right]}{\partial k_x} \right)\left( \frac{\partial \left[k'_yu(k')\right]}{\partial k'_x} \right)\int d\mathbf{r}_1  d\mathbf{r}_2 d\mathbf{r}_3 d\mathbf{r}_4 e^{i\mathbf{k}\cdot ( \mathbf{r}_1-\mathbf{r}_2)}e^{i\mathbf{k}'\cdot ( \mathbf{r}_3-\mathbf{r}_4)} \nonumber  \\
& &\times \Big\langle n(\mathbf{r}_1,t+\tau)n(\mathbf{r}_2,t+\tau)n(\mathbf{r}_3,t)n(\mathbf{r}_4,t)\Big\rangle, 
\label{deltaeta2}
\end{eqnarray}
which can also be written as 
\begin{eqnarray}
\Delta \eta (\tau;t)&=& \frac{(\beta/V) }{4(2\pi)^6}\int d\mathbf{k}\int d\mathbf{k}' \left( \frac{\partial \left[k_yu(k)\right]}{\partial k_x} \right)\left( \frac{\partial \left[k'_yu(k')\right]}{\partial k'_x} \right) \Big\langle n(\mathbf{k},t+\tau)n(-\mathbf{k},t+\tau)n(\mathbf{k}',t)n(-\mathbf{k}',t)\Big\rangle, \nonumber  \\
\label{deltaeta21}
\end{eqnarray}
where $n(\mathbf{k},t)$ is the Fourier transform $n(\mathbf{k},t)\equiv \int d\mathbf{r}\ e^{i\mathbf{k}\cdot \mathbf{r}} n(\mathbf{r},t)$ of $n(\mathbf{r},t)$.

Eqs. (\ref{deltaeta2}) and (\ref{deltaeta21}) are just two alternative exact manners to rewrite Eq. (\ref{greenkubo0}) for $\Delta \eta (\tau;t)$, with $\sigma ^{xy}(t)$ given by Eq. (\ref{eq: sigma1}), as an integral involving the four-point correlation function $\Big\langle n(\mathbf{r}_1,t+\tau)n(\mathbf{r}_2,t+\tau)n(\mathbf{r}_3,t)n(\mathbf{r}_4,t)\Big\rangle$ or $\Big\langle n(\mathbf{k},t+\tau)n(-\mathbf{k},t+\tau)n(\mathbf{k}',t)n(-\mathbf{k}',t)\Big\rangle$. It is important to notice that, so far, we have only assumed obvious and elementary  mathematical properties of the non-equilibrium average denoted by  the brackets $\langle ... \rangle$, but have never appealed to the condition of thermodynamic equilibrium. Although we did assume instantaneous thermalization, i.e., the condition of instantaneous \emph{thermal} equilibrium $T(\mathbf{r};t)=T^R(t)$, the degrees of freedom we are interested on in this work, are the configurational ones, whose relaxation is, of course, far slower and are the reason for the eventual dynamic arrest of the system. This, in fact, is what prevents us from identifying $\langle ... \rangle$ with an ordinary thermodynamic equilibrium average. The following steps are also general in this respect, but will no longer be exact, as we now discuss.

Beyond this point we have to introduce simplifying approximations, and the first of them is the Gaussian factorization, which approximates the four-point correlation functions referred to above, by a sum of products of two-point correlations. As shown in detail in Ref. \cite{wertheimlovett}, for example, within this approximation one can rewrite Eq. (\ref{deltaeta21}) as 
\begin{eqnarray}
\Delta \eta (\tau;t)&=& \frac{\beta n^2}{2(2\pi)^3}\int d\mathbf{k}\int d\mathbf{k}' \left( \frac{\partial \left[k_yu(k)\right]}{\partial k_x} \right)\left( \frac{\partial \left[k'_yu(k')\right]}{\partial k'_x} \right) \nonumber  \\
& & \times \Big\lbrace F(k,\tau;t)F(k',\tau;t)\delta(\mathbf{k}+\mathbf{k}') \Big\rbrace.   
\label{deltaetaApp3}
\end{eqnarray}
In addition, in the same reference the so-called Wertheim-Lovett relation of the equilibrium theory of inhomogeneous liquids \cite{evans} was also extended to non-equilibrium conditions leading, in particular, to the following approximate relation, 
\begin{eqnarray}
[k_y u(k)]= - k_BT S^{-1}(k;t)[k_y h(k;t)], 
\label{wlrelation0}
\end{eqnarray}
where $h(k;t)\equiv [S(k;t)-1]/n$ is the Fourier transform of the non-equilibrium total correlation function $h(r;t)$, and  $S(k;t)$ the non-equilibrium structure factor. Using now this relation, one can rewrite  Eq. (\ref{deltaetaApp3}) as 
\begin{eqnarray}
\label{deltaeta}
\Delta \eta (\tau;t)&=& \frac{k_BT}{60\pi^2}  \int_0^\infty dk k^4   \left[ \frac{1}{S(k;t)} \left(\frac{d S(k;t)}{d k} \right) \right]^2   \left[ \frac{F(k,\tau;t)}{S(k;t)} \right] ^2.
\end{eqnarray}
While this expression may initially appear identical to the one derived for thermodynamic equilibrium conditions, by Geszti   \cite{geszti}, and  by  N\"agele and Bergenholtz \cite{naegele,banchio}, it should be noted that it now expresses the time-dependent property $\eta(\tau;t)$ in terms of the non-equilibrium functions $S(k; t)$ and $F(k,\tau; t)$. Eq. (\ref{deltaeta}), complemented with the NESCGLE equations which we now discuss, constitutes the theoretical framework proposed in this work for the first-principles description of the non-equilibrium linear viscoelasticity.

\subsection{The  NE-SCGLE equations}\label{subsection2.1}

To calculate the stress relaxation function $\Delta\eta(\tau;t)$ from Eq. \eqref{deltaeta}, one requires the external determination of both, the time dependent structure factor $S(k;t)$ and non-equilibrium intermediate scattering function $F(k;\tau;t)$. As just mentioned, we can employ for this purpose the NE-SCGLE theory. The essential rationale and arguments to build this theoretical framework were laid down in Refs. \cite{nescgle1,nescgle2,nescgle3}, and a practical summary can be found in the supplemental material of Ref. \cite{zepeda}, which highlights the main simplifying approximations leading to the version of the NE-SCGLE theory employed here. More recently, however, the fundamental principles upon which this theory is based, were explained in a more formal manner by identifying which of these principles are genuine \emph{physical} principles, and which are in reality restrictions imposed by the mathematical framework employed to describe these physical principles and their implications \cite{nonlinonsmach,wertheimlovett}. 

The simple version of the NE-SCGLE theory employed here can be summarized by a set of equations describing the time-evolution of the non-equilibrium structural and dynamical properties of a model colloidal liquid in the absence of hydrodynamic interactions, comprised by $N$ identical particles in a volume $V$ that diffuse with a free-diffusion coefficient $D^0$ while interacting through radially-symmetric forces whose pair potential is $u(r)$. The core of the NE-SCGLE theory is an equation of motion describing the time evolution of $S(k;t)$. When the liquid is instantaneously quenched (at time $t=0$) from initial bulk density and temperature ($n_i$,$T_i$) to new final values ($n$,$T$), and is constrained to remain spatially uniform, such equation reads (for $t>0$) 
\begin{equation}
\frac{\partial S(k;t)}{\partial t} = -2k^2 D^0 b(t)n\mathcal{E}(k) \left[S(k;t)-1/n\mathcal{E}(k)\right].  \label{relsigmadif2appendix}
\end{equation}
As carefully explained in Refs. \cite{nescgle1,nescgle2,nescgle3}, the thermodynamic stability function $\mathcal{E}(k)$ is the Fourier transform (FT) of the functional derivative $\mathcal{E}[\mid \mathbf{r}-\mathbf{r}'\mid ; n,T] \equiv  \left[ \delta^2 (\mathcal{F}[n,T]/k_BT)/\delta n(\mathbf{\mathbf{r}})\delta n(\mathbf{\mathbf{r}'}) \right]$ of the Helmholtz free energy density-functional $\mathcal{F}[n;T]$, evaluated at the final state point $(n,T)$. 

In addition, the mobility function $b(t)$ is defined as $b(t)\equiv D_L(t)/D^0$, with $D_L(t)$ being the long-time self-diffusion coefficient at evolution time $t$. This function couples the relaxation of $S(k;t)$ with the non-equilibrium relaxation of the dynamical properties of the liquid. More specifically, such coupling is established by the following exact expression for $b(t)$,
\begin{equation}
b(t)= \Big[1+\int_0^{\infty} d\tau\Delta{\zeta}^*(\tau; t)\Big]^{-1},
\label{bdt}
\end{equation}
in terms of the $t$-evolving and $\tau$-dependent friction function $\Delta{\zeta}^*(\tau; t)$, for which the NE-SCGLE theory also provides the approximate result 
\begin{equation}
  \Delta \zeta^* (\tau; t)= \frac{D_0}{24 \pi^{3}n}\int d {\bf k}\ k^2 \left[\frac{ S(k;t)-1}{S(k; t)}\right]^2  F(k,\tau; t)F_S(k,\tau; t),
\label{dzdtquench}
\end{equation}
written in terms of $S(k; t)$, $F(k;t)$ and the self ISF $F_S(k,z; t)=\langle \exp[i\mathbf{k}\cdot \Delta\mathbf{r}_T(\tau;t)]\rangle$, with $\Delta\mathbf{r}_T(\tau;t)\equiv\mathbf{r}_T(t+\tau)-\mathbf{r}_T(t)$ being the displacement of one of the $N$ identical particles over a time interval $\tau$. The NE-SCGLE also provides time-evolution equations for the above non-equilibrium dynamic correlations. In terms of their Laplace transforms, $F(k,z; t)$ and $F_S(k,\tau; t)$, such equations read
\begin{gather}\label{fluctquench}
 F(k,z; t) = \frac{S(k; t)}{z+\displaystyle{\frac{k^2D^0 S^{-1}(k;
t)}{1+\lambda (k)\ \Delta \zeta^*(z; t)}}},
\end{gather}
and
\begin{gather}\label{fluctsquench}
 F_S(k,z; t) = \frac{1}{z+\displaystyle{\frac{k^2D^0 }{1+\lambda (k)\ \Delta
\zeta^*(z; t)}}},
\end{gather}
with $\lambda (k)$ being a phenomenological interpolating function, given by
\begin{equation}
\lambda (k)\equiv 1/[1+( k/k_{c}) ^{2}],
\label{lambdadk}
\end{equation}
and where $k_c$ is an empirically cutoff wave-vector. Within the NE-SCGLE, this parameter is employed to \emph{calibrate} the theory for each specific application (see for instance Sec. III of Ref. \cite{mendoza}).

Thus, for a given system and protocol of fabrication (i.e. for given pair potential $u(r)$ and final state point $(n,T)$), the solution of Eqs. \eqref{relsigmadif2appendix}-\eqref{lambdadk} provides the $t$-evolution of the functions $S(k;t)$, $b(t)$, $\Delta\zeta^*(\tau;t)$, $F(k,\tau;t)$ and $F_S(k,\tau;t)$, which describe the irreversible relaxation of an instantaneously and homogeneously quenched liquid. The specific details regarding the numerical solution of Eqs. \eqref{relsigmadif2appendix}-\eqref{lambdadk} can be found in Ref. \cite{nescgle3}.

\section{Illustrative application}\label{section3}

As said in the introduction, combining the expression in Eq. (\ref{deltaeta}), which writes the total shear-stress relaxation function $\eta(\tau;t)=2\delta(\tau)\eta^0+\Delta\eta(\tau;t)$ in terms of $S(k; t)$ and $F(k,\tau; t)$, with the solution of the NE-SCGLE equations for  these non-equilibrium properties (Eqs. \eqref{relsigmadif2appendix}-\eqref{lambdadk}), constitutes a general protocol for the first-principles description of the aging of the viscoelastic properties of a fluid instantaneously quenched into a dynamically-arrested  (i.e., glass or gel) state. Such general protocol provides the non-equilibrium evolution of $\eta(\tau;t)$ itself and, of course, of the other physical observables that derive from this property, such as the instantaneous total viscosity $\eta(t)\equiv \int _0^\infty \eta(\tau;t) d \tau$, the dynamic  shear viscosity $\eta(\omega;t)$, or the elastic and loss moduli, $G'(\omega;t)$ and $G''(\omega;t)$.

Thus, in this section we provide an illustrative example of the application of such protocol. For this, we shall focus on a specific glass-forming model system, namely, the Weeks-Chandler-Andersen (WCA) fluid \cite{wca}, for which the NE-SCGLE theory provides an accurate description of the irreversible relaxation of the correlation functions $S(k;t)$ and $F(k,\tau;t)$ after an instantaneous quench in control parameters \cite{gabriel,nescgle5,mendoza,rivas}. Let us recall that the WCA potential vanishes for $r$ larger than the distance $\sigma$ of soft contact, but for $r\leq\sigma$ is given by
\begin{equation}
 u(r) =
\epsilon\left[ \left( \frac{\sigma}{r}\right)^{12}
-2\left( \frac{\sigma}{r}\right)^{6}+1 \right].
\label{truncatedlj}
\end{equation}
The state space of this system is spanned by the volume fraction $\phi = \pi n \sigma^3/6$ and the effective temperature $T^*\equiv k_BT /\epsilon$. For simplicity, from now on we will refer to the dimensionless parameter $T^*$ just as $T$, and use $\sigma$ and [$\sigma^2/D^0$] as the units of length and time, respectively.

In order to solve the NE-SCGLE equations \eqref{relsigmadif2appendix}-\eqref{lambdadk}, one requires the external determination of the fundamental thermodynamic input $\mathcal{E}(k;n,T)$ involved in Eq. \eqref{relsigmadif2appendix}, which is the Fourier transform of the functional derivative $\mathcal{E}[|\mathbf{r}-\mathbf{r}'|;n,T]\equiv\delta\beta\mu[\mathbf{r};n]/\delta n(\mathbf{r}')$ of the chemical potential $\mu[\mathbf{r};n]$ with respect to the local density $n(\mathbf{r}')$, evaluated at the uniform concentration and temperature profiles, $n(\mathbf{r})=n$ and $T(\mathbf{r})=T$, of the final state of a quench. 
In turn, $\mathcal{E}(k;n,T)$ is related to the FT of the two-particle direct correlation function $c[|\mathbf{r}-\mathbf{r}'|;n;T]$, and to the  equilibrium structure factor $S^{eq}(k;n,T)$ by $n\mathcal{E}(k;n,T)=1-n c(k;n,T)=1/S^{eq}(k;n,T)$, with $n$ being the macroscopic density, $n=N/V$, and $\beta^{-1}\equiv k_BT$ \cite{wertheimlovett}. As also discussed  in detail in Refs. \cite{nescgle3,mendoza}, in our present case this property can be determined using the Ornstein-Zernike (OZ) equation combined with the Percus-Yevick/Verlet-Weiss (PY-VW) approximation \cite{percus,verlet} for an effective HS fluid with a temperature dependent HS diameter $\sigma(T)$ obtained by the blip function 
method \cite{hansen} (see also Sec. IIIA of Ref. \cite{nescgle3}). 

\subsection{Non-equilibrium viscoelastic behavior of the WCA liquid}\label{subsection3.1}

Let us first discuss the non-equilibrium viscoelastic behavior of the WCA liquid  predicted by Eq. \eqref{deltaeta} to occur after an isochoric quench. In fact, for methodological convenience we will describe two illustrative independent quenches, both starting from the same ergodic state point ($\phi=0.7, T_i=1$) but which consider different final temperatures, namely, $T=0.25$ and $T=0.1$, which lie, respectively, above and below the dynamic-arrest temperature $T^a(\phi=0.7)=0.235$  of that isochore (see the inset of Fig. \ref{fig:1}(a)). These two quenches are representative, respectively, of the transient process of equilibration, and of the ultra-slow process of aging in the WCA fluid, as discussed in Ref. \cite{mendoza}, whose Secs. III and IV provide the details of the relaxation of the functions $S(k;t)$ and $F(k,\tau;t)$ along these two processes. The present work, in contrast, illustrates the most salient out-of-equilibrium characteristics predicted by Eq. \eqref{deltaeta} for the stress relaxation function $\Delta\eta(\tau;t)$, as well as for the corresponding instantaneous viscosity $\eta(t)$, and the elastic and loss moduli $G'(\omega;t)$ and $G''(\omega,t)$. To be precise, in what follows we shall refer to the normalized  total shear-stress relaxation function $\eta^*(\tau;t)\equiv \eta(\tau;t)/\eta^0= 2 \delta (\tau) + \Delta\eta^*(\tau;t)$, with $\Delta\eta^*(\tau;t)\equiv \Delta\eta(\tau;t)/\eta^0$ being the contribution of the direct interactions between particles. The normalizing factor $\eta^0$ is the short-time  viscosity. Integrated on $\tau$, these two properties become the normalized instantaneous value of, respectively, the ordinary viscosity  $\eta^*(t)=1+ \Delta\eta^*(t) $ and stress relaxation function $\Delta\eta^*(t) $.


\subsubsection{Equilibration of $\eta^*(t)$ and $\Delta\eta^*(\tau;t)$}\label{subsection3.2}

\begin{figure}
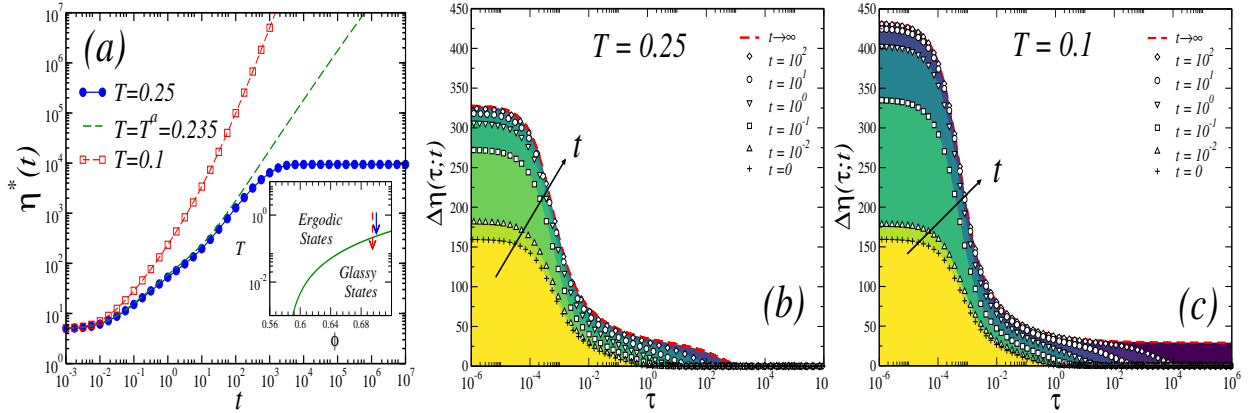
 
\includegraphics[width=0.33\textwidth, height=0.33\textwidth]
{eta.eps}
\includegraphics[width=0.33\textwidth, height=0.33\textwidth]
{Viscosity0p25.eps}\includegraphics[width=0.33\textwidth, height=0.33\textwidth]
{Viscosity0p1.eps}  
\caption{(a) Instantaneous viscosity $\eta^*(t)\equiv\eta(t)/\eta^0$ as a function of the waiting time $t>0$ of the WCA fluid,  initially at equilibrium ($t<0$) at the state point ($\phi=0.7,T_i=1$), but which at $t=0$ was isochorically and instantaneously quenched to a final temperature $T=0.25$ (still in the ergodic region; solid circles), and to a final temperature $T=0.1$ (well in the glassy region; open squares). Inset, dynamical arrest diagram of the WCA predicted by the NE-SCGLE theory and reported in Ref. \cite{mendoza}. The two vertical arrows highlight the two isochoric quenches above ($T=0.25$) and below ($T=0.1$) the dynamic arrest temperature $T^a(\phi=0.7)=0.235$, thus corresponding to an equilibration and an aging process, respectively (see the text).  (b) and (c) Sequence of snapshots illustrating the waiting time evolution of the relaxation function $\Delta\eta(\tau;t)$, plotted as a function of the delay time $\tau$,  for the two isochoric quenches displayed in (a). A gradation of colors (ranging from yellow to blue) is used to highlight the temporal evolution of the area under the curve of the function $\Delta\eta^{*}(\tau;t)$ (see the text).}\label{fig:1} 
\end{figure}

Fig. \ref{fig:1}(a) displays the time evolution of $\eta^*(t)$ obtained from Eq. \eqref{deltaeta} for the two aforementioned processes. For the shallower quench, $T=0.25$ (solid circles), the viscoelastic response of the system to an instantaneous cooling  consists of a transient development of $\eta^*(t)$ with waiting time, which grows sublinearly from its initial equilibrium value, $\eta^{*eq}(\phi=0.7,T_i=1)=\eta^*(t=0)$, over a time window of roughly three decades. Beyond this time scale, the instantaneous viscosity acquires a new constant value, $\eta^*_f$, thus indicating the full equilibration of the system. Notice that, along this transient process, the viscosity of the WCA liquid is predicted to increase approximately three orders of magnitude. 

Since the instantaneous viscosity is written as $\eta^*(t) =1+ \Delta \eta^*(t)$, with $\Delta \eta^*(t) \equiv \int_0^\infty d \tau \Delta\eta^*(\tau;t)$, we have that $\Delta \eta^*(t)$ is actually the area under the function $\Delta\eta^*(\tau;t)$ plotted as a function of the delay time $\tau$. It is then instructive to display the process of equilibration, exhibited  as the increase and saturation of $\eta^*(t)$ in Fig. \ref{fig:1}(a), now  in terms of the time evolution of the  function $\Delta \eta^*(\tau;t)$. This is done in Fig. \ref{fig:1}(b), which illustrates the equilibration process of the WCA system to the final temperature $T=0.25$, in terms of the evolution of the $\tau$-dependence of $\Delta\eta^*(\tau;t)$ (open symbols along the borderline between colors), at a sequence of values of the waiting time $t$. 

One notices in Fig. \ref{fig:1}(b) that as $t$ becomes larger, the function $\Delta\eta^*(\tau;t)$ increases for all $\tau$, thus leading to a gradual growth of the area under its curve (i.e., to an increase in $\eta^*(t)$). For reference, such area coverage is highlighted in Fig. \ref{fig:1}(b) using different colors (from yellow to blue). In addition, for waiting-times of order $10^3$ and longer, $\Delta\eta^*(\tau;t)$ no longer evolves, but saturates to a stationary profile $\Delta\eta^{*}_f(\tau)$,  which coincides with the asymptotic equilibrium value $\Delta\eta^*(\tau;t\to\infty)\equiv\Delta\eta^{*eq}(\tau)$ (dashed line) obtained from Eq. \eqref{deltaeta} using as inputs the stationary equilibrium solutions $S(k;t\to\infty)=S^{eq}(k)$ and  $F(k,\tau;t\to\infty)=F^{eq}(k;\tau)$ of the NE-SCGLE equations \eqref{relsigmadif2appendix}-\eqref{lambdadk} (see, for reference, Sec. III of Ref. \cite{mendoza}). The $\tau$-dependence of the final equilibrium value $\Delta\eta^{*eq}(\tau)$ (dashed line) clearly exhibits an early fast decay, associated with the so-called $\beta$-relaxation processes, followed by an inflection point at $\tau \approx 10^{-1}$, which starts an incipient plateau  that lasts for about two decades, ending with a much slower final decay to zero, associated with the (configurational) $\alpha$-relaxation processes. The full sequence of snapshots of $\Delta\eta^*(\tau;t)$ in Fig. \ref{fig:1}(b) exhibits the build-up of this two-step relaxation scenario, starting from the initial value $\Delta\eta^*(\tau;t=0)$, which essentially exhibits only the initial $\beta$-relaxation.

\subsubsection{Aging of $\eta^*(t)$ and $\Delta\eta^*(\tau;t)$}\label{subsection3.3}

The above features regarding the evolution of the instantaneous viscosity the WCA system during an equilibration process are to be contrasted against those obtained for a deeper quench into the glass regime. Coming back to Fig. \ref{fig:1}(a), for example, one notices a faster increase in $\eta^*(t)$ for a quench to $T=0.1$ (open squares), in reference to that observed for $T=0.25$ at comparable waiting times (approximately two orders of magnitude larger at $t=10^2$). More importantly, however, in this case one finds a persistent growth of $\eta^*(t)$, at all $t$ and without any sign of saturation, thus implying an endless aging process into the glass state, in which the viscosity, while remaining finite, continues to increase as the system ages. Thus, the two predicted patterns in Fig. \ref{fig:1}(a) illustrate the kinetics of aging ($T<T^a(\phi)$) and of equilibration ($T>T^a(\phi)$), with a crossover from one regime to the other occurring exactly at the arrest temperature $T^a(\phi)=0.235$ (dashed line). 

Overall, we found that the predicted behavior of $\eta^*(t)$ turns out to be quite similar to that reported in Fig. 4(a) of Ref. \cite{mendoza}, regarding the $t$-evolution of the $\alpha$-relaxation time $\tau_\alpha(t)$ (defined by the condition $F(k_{max},\tau_\alpha(t);t)=1/e$, with $k_{max}$ being the position of the main peak of $S(k;t)$). In fact, that figure exhibited the excellent qualitative and quantitative agreement of the predicted crossover scenario for $\tau_\alpha(t)$, with the results of Brownian dynamics simulations at short and intermediate times. As discussed in Subsection IV.D of Ref. \cite{mendoza}, only at long times, and only for the aging processes with $T\leq T^a$, noticeable deviations were observed, probably due to the spatial heterogeneity of the local density, unavoidably present in microscopic simulations and in real experiments, but not yet included in the present theoretical calculations \cite{elizondo_prl}. Here we conjecture that a similar scenario will be observed when the present predictions for the viscosity $\eta^*(t)$ are contrasted with simulations or experiments.

Just like in the equilibration process, the above aging behavior can also be visualized in the $t$-evolution of the stress relaxation function $\Delta\eta^*(\tau;t)$, shown for completeness in Fig. \ref{fig:1}(c). Focusing on the regime of small correlation times ($\tau\leq10^{-2}$, i.e.,  the $\beta$-relaxation regime), one sees that $\Delta\eta^*(\tau;t)$ displays in general higher values, with respect to the case $T_f=0.25$, thus leading to increasingly larger contributions to $\eta^*(t)$. For waiting times of order $10^{2}$, however, the stress relaxation function at small $\tau$ exhibits saturation. In contrast, the long-tailed relaxation of $\Delta\eta^*(\tau;t)$ (for $\tau \gtrsim 10^2$) becomes gradually slower as the system ages and, more importantly, does not show any sign of saturation to a stationary value. Instead, this long-$\tau$ tail starts at intermediate waiting times ($t \approx 10^2$) by exhibiting an incipient plateau, which becomes much better defined at longer waiting times (see, for example, the result for $t =10^4$). This evolving plateau coincides increasingly better with the non-decaying plateau of the  asymptotic solution $\Delta\eta^{*a}(\tau)\equiv\Delta\eta^*(\tau;t\to\infty)$ (red dashed line), obtained from Eq. \eqref{deltaeta} using the asymptotic arrested solutions $S^a(k)$ and $F^a(k,\tau)$ of the NE-SCGLE equations for the quench with $T=0.1$ (see Sec. IV of Ref. \cite{mendoza}).   As illustrated by the results in Fig. \ref{fig:1}(c), the longest the waiting time, the longest the stretch in which this transient plateau coincides with the arrested plateau of $\Delta\eta^{*a}(\tau)$. However, for any finite $t$, the transient plateau will always end in the decay of $\Delta\eta^{*}(\tau;t)$ to zero with $\tau$, leading to a finite value of $\eta^*(t)$.

\subsubsection{Dynamical Shear Modulus}

\begin{figure*}[ht]
        \centering
\includegraphics[width=0.32\textwidth, height=0.3\textwidth]{Fig2a_new_full.eps}
\includegraphics[width=0.32\textwidth, height=0.3\textwidth]{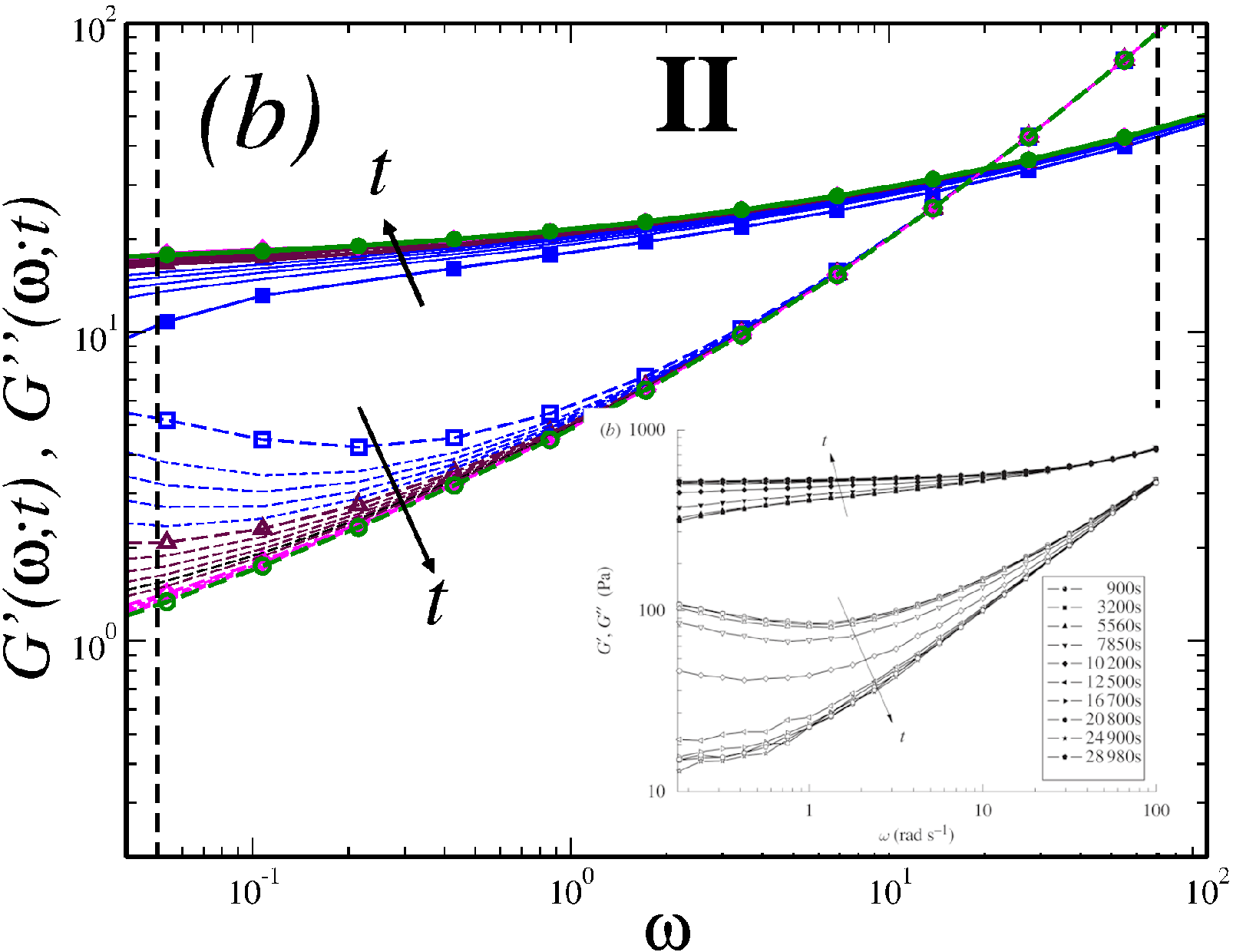}
\includegraphics[width=0.32\textwidth, height=0.3\textwidth]{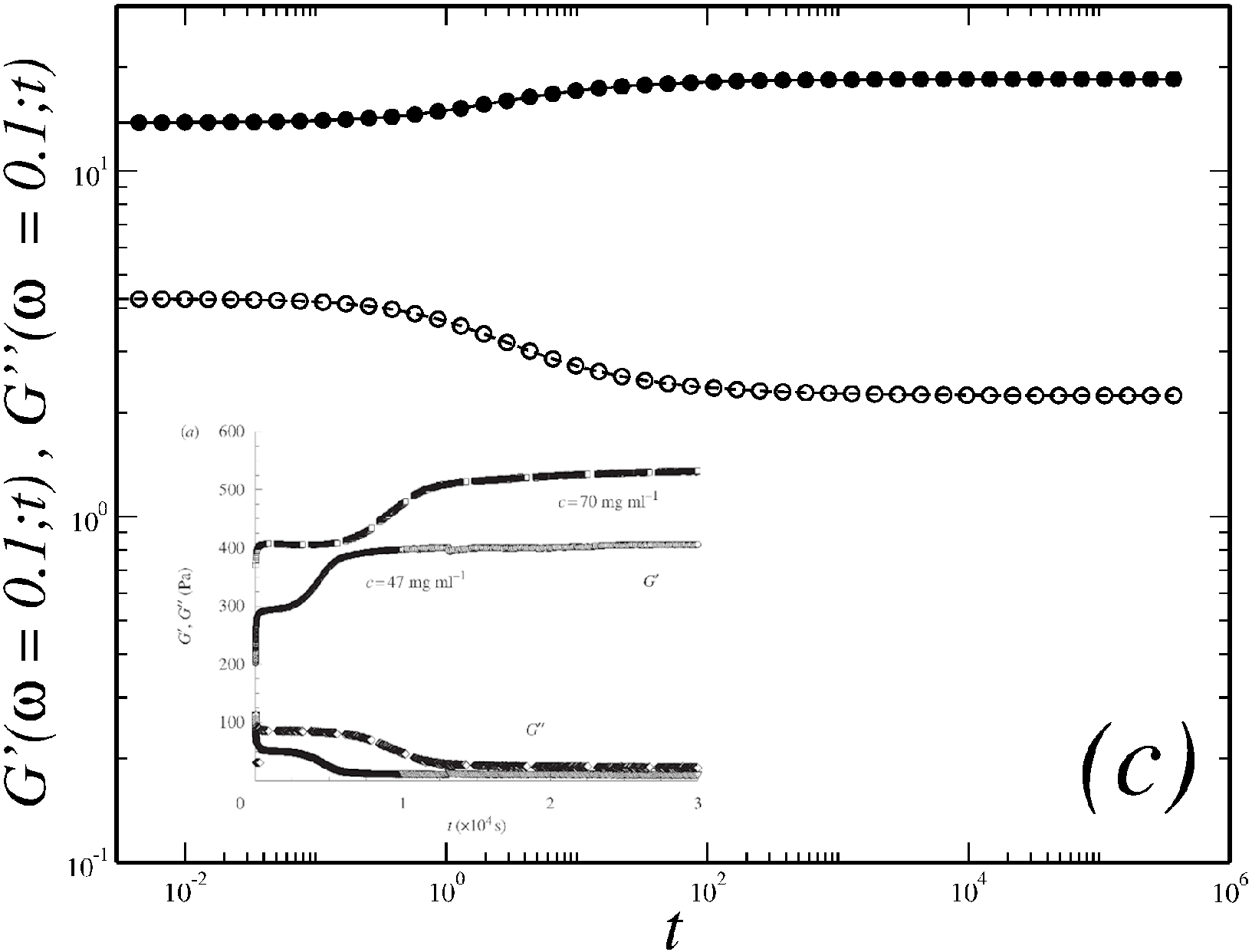}
        \caption{  (a) Snapshots of the theoretically predicted dynamic moduli $G'(\omega;t)$ (solid symbols) and $G''(\omega;t)$ (open symbols), plotted as functions of $\omega$ for a sequence of waiting times $t$ (symbols and lines), for the same quench of the WCA model as in Fig. 1(b) and exhibiting three regimes at (I) low, (II) intermediate and (III) high frequencies. (b) Same as in (a), but zooming into the regime II to qualitatively compare with the experimental data shown in the inset, taken from Fig. 3(b) of Ref. \cite{christopoulou}. (c) Time evolution of $G'(\omega;t)$ and $G''(\omega;t)$ at fixed $\omega=0.1$. The inset shows the experimental data reported in Fig. 3(a) of Ref. \cite{christopoulou}.}
        \label{fig: TheoryVsExp}
    \end{figure*}

Other viscoelastic properties of practical and fundamental interest directly derive from the relaxation function $\eta(\tau;t)$. For example, the aging process illustrated with the evolution of the instantaneous viscosity $\eta(t)$ in Fig. 1(a), can also be displayed by plotting the $\omega$-dependence of the storage and loss moduli $G'(\omega;t)$ and $G''(\omega; t)$ at a sequence of waiting times. The continuous evolution of these rheological properties with waiting time is another distinctive signature of glassy and gelled states of matter, as has been shown in various systematic experimental characterizations 
\cite{purnomo,purnomo1,christopoulou,suman}. Hence, this alternative format is now presented in Fig. \ref{fig: TheoryVsExp}(a), which displays the non-equilibrium evolution of $G'(\omega;t)$ and $G''(\omega; t)$ as predicted by Eq. \eqref{deltaeta} for the same aging process, $T=0.1$. Notice that the sequence of snapshots in this figure now correspond to $t=10^1,10^2,10^3$ and $10^4$ (symbols), and to additional (arbitrary) times in between (thin lines). 

In the wide frequency range of this figure, we identify three distinct domains for the qualitative behavior of these two moduli, corresponding to (I) low, (II) intermediate (or "experimental") and (III) high frequencies, thus describing, respectively, the long-, intermediate-, and short-time relaxation of the system. Since aging is essentially a slow process, its effects are most noticeable at small frequencies (domain I), whereas they are virtually absent at short times (domain III).  Thus, it is mostly in the crossover regime (domain II) that the aging effects can be measured within practical waiting times \cite{purnomo,purnomo1,christopoulou,suman}. For this reason, in Fig. 2(b) we present an amplification of  Fig. 2(a), centered on this crossover regime, to better appreciate its qualitative similarity with experimental measurements.

As seen in Fig. 2(b), in this region $G'(\omega;t)$ (solid symbols) is predicted to become progressively larger with $t$, whereas $G''(\omega;t)$ (open symbols) follows the opposite trend and evolves towards smaller values, with both quantities gradually approaching an asymptotic value at large waiting times. One notices, however, that the waiting time needed to reach this asymptotic value is significantly longer for $G''(\omega;t)$, which thus contains the most noticeable aging effects. Let us highlight that this predicted physical scenario is strikingly similar to the experimental data of the aging of the dynamic moduli of a suspension of purely repulsive star polymers in a glassy state, reproduced for clarity in the inset of Fig. 2(b), from the work of  Christopoulou et al. (Fig. 3 of Ref. \cite{christopoulou}). These data clearly exhibit the same trends of the main figure despite the obvious difference between the WCA model pair potential $u(r)$ of Eq. (4) and the effective soft potential $u_{SP}(r)$ between two star polymers (see, for example, Eq. (39) of Ref. \cite{06Likos}). It is worth mentioning that very similar trends have also been reported for dense colloidal suspensions of thermosensitive particles \cite{purnomo,purnomo1}. We conjecture that the origin of these similarities is the possible universality of the principle of structural and dynamical equivalence between soft-sphere fluids of different  softness, well-documented under equilibrium conditions \cite{dyneq0,dyneqPRL,dyneqPRE1,dyneqPRE2}. Within well-defined scalings, this equivalence includes the star polymer experimental system, the WCA model, and the hard-sphere limit. It will, of course, be interesting to see to what extent the notion of hard-sphere dynamic universality class \cite{dyneqPRL} might extend over to the realm of non-equilibrium phenomena. We leave such exercise for further work.

A complementary format to describe the aging of the viscoelastic response is to consider a dynamic time sweep of the storage and loss moduli at fixed $\omega$ \cite{suman,purnomo,purnomo1}. Such information is displayed in Fig. 2(c) for $G'(\omega=0.1;t)$ (solid symbols) and $G''(\omega=0.1;t)$ (open symbols), plotted now as a function of $t$. Different from Figs 1(a) and 1(b), however, in Fig. 2(c) we have considered a different quench, now starting from $T_i=0.2$ but ending again at $T=0.1$, thus resembling more closely the experimental conditions of Ref. \cite{christopoulou}. At short $t$, both moduli describe a transient plateau, which precedes a very slow but persistent evolution of $G'(\omega=0.1;t)$ and $G''(\omega=0.1;t)$, in which the former becomes gradually larger, while the later becomes increasingly smaller. Again, these trends resemble qualitatively the data of Ref. \cite{christopoulou}, reproduced for reference in the inset of Fig 2(c).

Let us emphasize, however, that the intention of the comparisons above, is only to illustrate the predictive capacity of the approach proposed in this work to the first-principles theoretical modeling of the aging of the linear viscoelastic properties of glass- and gel-forming liquids. A systematic and rigorous comparison of the results of the proposed theoretical scheme with its experimental counterpart is left as a future exercise which shall be addressed elsewhere.

\subsection{Meaningful consequences}\label{subsection3.4}

The fact that the relaxation function $\Delta\eta^*(\tau;t)$ will always decay to zero with $\tau$ at any experimentally accessible (i.e., finite) waiting time, even for $T\leq T^a$, is a meaningful prediction of the NE-SCGLE theory. It means that any real experiment (performed, of course, at finite $t$) will report an increasingly larger (\emph{but always finite!}) viscosity $\eta(t)$. Observing a divergent viscosity would require an infinite waiting time, an idealized physical scenario impossible to be observed in practice. As carefully explained in Refs. \cite{nescgle3,mendoza}, this more realistic scenario for the precipitous growth of the relaxation times and of the viscosity in a glass forming-liquid, complements and clarifies some fundamental aspects of the notion of ``ideal glass transition" predicted by equilibrium theories, such as mode coupling theory (MCT), which only describe the stationary limit $t=\infty$. The kinetic extension to finite waiting times ($0\le t < \infty$) provided by the NE-SCGLE predictions, extended here to the level of the non-equilibrium viscoelastic response of a glass forming system illustrated in Fig. \ref{fig:1}, opens the possibility of a more meaningful and consistent comparison of theoretical, simulated and experimental results. 
 
A few additional remarks are in order here. In spite of the lack of consensus for its first-principles explanation, it is generally agreed that the experimental phenomenon referred to as ``the glass transition" has a kinetic origin. Indeed, when a real glass-forming liquid is quenched into its glassy regime, the time for its structural relaxation exceeds the experimentally accessible times and, consequently, a slow and long lasting evolution of the structural, dynamical and rheological properties of the liquid is observed. The fundamental understanding of the resulting phenomenology of aging and dynamical arrest -- which is characteristic of a large variety of gelled and glassy materials with very different microscopic characteristics -- is severely limited by the finite time-span of any practical observation, either experimental or simulated.  While most experimentalists are fully familiar with the empirical fact that a large increase in the relaxation time comes with a concomitant increment of the liquid's viscosity, the specific way in which these two physical observables are connected is generally unknown. Determining such connection, in fact, remains as one of the most challenging problems for theoretical physics and materials science. It is thus very important to highlight that Eq. \eqref{deltaeta} constitutes an innovative proposal to establish such connection and, more crucially, to provide a direct route to relate both, the kinetics of the structural relaxation and the viscoelastic response, with intermolecular forces and fabrication protocols. To the best of our knowledge, this is the first time that such first-principles link is established.

\section{Time-temperature-density diagrams}\label{section4}

In the previous section we have presented the description provided by the NE-SCGLE theory, of the viscoelastic properties of a glass-forming model liquid.  To distinguish between two different (and mutually exclusive) physical scenarios, equilibration and aging, we restricted ourselves to consider only two representative individual quench processes. Under many circumstances, however, the interesting concept is not the rheological behavior of a single system after one individual quench, but the collective scenario revealed by an ensemble of many representative independent systems subjected to many different and independent quench processes. In this case, the main goal is to learn about the collective evolution of the ensemble in the multi-dimensional space of relevant physical properties and  control parameters. In this section we discus two illustrative examples of this possibility.

There may be many ways of displaying and analyzing the information encoded in the dependence of the total shear stress relaxation function $\eta(\tau;t)$ on the correlation or delay time $\tau$, and on the waiting time $t$. As just discussed, this includes considering the instantaneous value of the viscosity $\eta(t)$ or, alternatively, the $\omega$ and $t$ dependence of the dynamic shear viscosity $\eta(\omega;t)$, leading to the elastic and loss moduli, $G'(\omega;t)$ and $G''(\omega;t)$.  Let us recall, in addition, that any of these properties also depend on the protocol of preparation, represented in our case by the initial  and final state points of the quench, $(\phi_i,T_i)$ and $(\phi,T)$. Then, displaying and analyzing a function of so many variables, such as $\eta(\tau;t;\phi_i,T_i,\phi,T)$, may become an extremely rich, but also overwhelming task, unless we reduce the number of independent variables by considering subspaces of the full 6-dimensional space $(\tau;t;\phi_i,T_i,\phi,T)$, starting with the experimentally most relevant of such subspaces. In this spirit, we have started in the previous section by integrating out the correlation time $\tau$, to consider only the instantaneous total viscosity $\eta(t)\equiv \int_0^\infty d \tau \eta(\tau;t)$, and by fixing the initial condition of the quench, represented by $(\phi_i,T_i)$. In this manner, we focused only on the description of the total viscosity, as the function $\eta(t;\phi,T)$ of only three variables (so far, restricted to the case $\phi=0.7$). 

The simplest manner to analyze the NE-SCGLE predictions for $\eta(t;\phi,T)$ is to relate them with real or hypothetical experimental contexts, and in this section we explain this notion with two examples. In both cases we describe the function $\eta(t;\phi,T)$ in terms of the predicted behavior of an ensemble of systems quenched at many different final state points, representative of a given subspace of the state space $(\phi,T)$. The subspace considered in the first example is an isochore line, where we fix the volume fraction to $\phi=0.7$ and display the dependence of the viscosity on the final temperature $T$ and the waiting time $t$. This information is the essence of the empirical concept referred to as \emph{Angell plot} \cite{angell}, enriched here by the kinetic perspective provided by the dependence on the waiting time $t$. The second example is a simple extension of this exercise, in which rather than fixing the volume fraction, we display the dependence of $\eta(t;\phi,T)$ along other subspaces, thus predicting the essential information corresponding to the empirical notions of time-temperature-transformation (TTT) diagram \cite{callisterrethwisch}  or time-density-transformation (TDT) diagram  \cite{ruzicka1}.

\subsection{Time-dependent non-equilibrium viscosity.}\label{subsection4.1}

The temperature dependence of  the logarithm of the zero shear-rate viscosity $\eta(T)$ of glass forming liquids, plotted versus $T_g /T$, is the essence of what are commonly referred to as \emph{Angell plots}, which are particularly useful for studying the behavior of materials near their glass transition \cite{angell}. In such  plots,  $T_g$ is the experimental glass transition temperature, empirically defined as the temperature at which $\eta(T)$ reaches the value $\eta(T_g) = 10^{12}$ Pa.Sec. (or $\eta^*(T)\equiv \eta(T)/\eta^0$ reaches the value $\eta^*(T_g) = 10^{16}$, since $\eta^0 \approx 10^{-4}$ Pa.Sec.). This $T$-dependence of the viscosity provides valuable insights on the experimental correlations with other relevant properties of structural glasses \cite{huangmckenna}, revealed by an enormous number of experimental and computer simulation studies \cite{kelton} and often successfully rationalized in terms of intuitive heuristic models \cite{mauroallan}. 

In contrast, in spite of a long and rich theoretical discussion \cite{berthierbiroli}, the possibility of theoretically explaining and predicting the observed $T$-dependence of the viscosity from physical \emph{first-principles} has been remarkably absent in the literature. In fact, this possibility has even been considered ``a dream'' \cite{tarjus}. This skepticism, however, has been confronted by MCT  (see \cite{jensen} and references therein) as well as by the equilibrium version of the NE-SCGLE theory \cite{todos1,reviewroque,eliz_voigt}. Although these two first-principles theories only describe the dynamics of equilibrium liquids, and predict a divergence of the viscosity that is never observed in practice, in what follows we demonstrate that the application of Eq. \eqref{deltaeta}, together with the NE-SCGLE equations in Subsection \ref{subsection2.1}, successfully remove these limitations. To illustrate this claim, in what follows we discuss a simple example involving again the WCA model.

\begin{figure}\includegraphics[width=0.5\textwidth, height=0.5\textwidth]{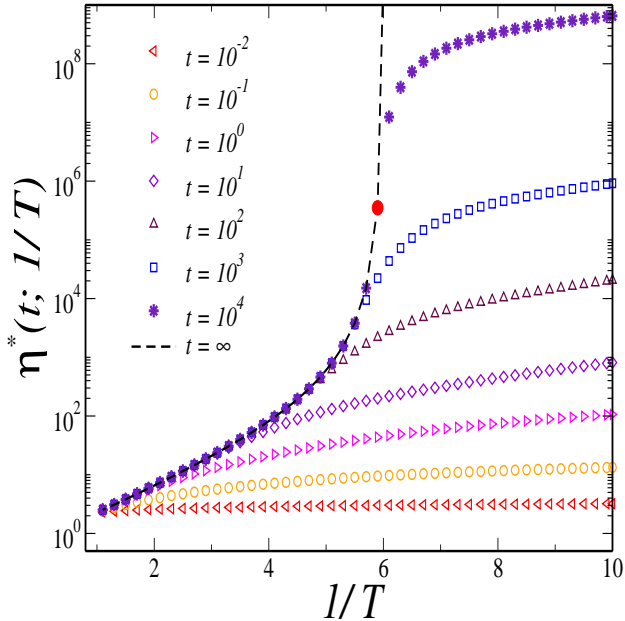}
\caption{Behavior of the function $\eta^*(t;\phi_i=0.7,T)$ for a sequence of isochoric quenches of the WCA system, starting from the equilibrium state point ($\phi=0.7;T_i=1$), and considering different final temperatures $T$. For clarity in the discussion, we plot $\eta^*(t)$ vs $1/T$ for different waiting times, as indicated (see the text)}.  \label{fig:2}
\end{figure}

Hence, let us apply now Eq. \eqref{deltaeta} and the NE-SCGLE equations to the determination, not only of the $T$-dependence of the viscosity of a glass forming liquid, but also of its dependence on the waiting  time $t$ at which one starts its measurement. This possibility of describing the aging of the viscoelastic properties derives from the powerful kinetic perspective provided by the NE-SCGLE, illustrated in the previous section only with two individual quenches. Let us now enrich that exercise by considering not two, but a larger set of independent representative systems, simultaneously subjected to an instantaneous isochoric quench at $t=0$, starting from the same initial equilibrium conditions at the state point $(\phi=0.7,T_i=1)$, but with different final temperature $T$. The set of values of $T$ will cover both regimes, above and below the dynamic arrest temperature $T^a(\phi)$ along the isochore $\phi=0.7$. 

Based on Eqs.  \eqref{deltaeta} and  \eqref{relsigmadif2appendix}-\eqref{lambdadk}, we have evaluated the total shear viscosity $\eta^*(t;\phi,T)$ for this ensemble of quenches at fixed  $\phi=0.7$. Fig. \ref{fig:2} plots the results for $\eta^*(t;1/T)\equiv \eta^*(t;\phi=0.7,T)$ as a function of the inverse temperature $1/T$, for the indicated waiting times. For reference, the same figure also includes the asymptotic value $\eta^*(t\to\infty;1/T)=\eta^{*eq}(1/T)$ (black dashed line), which coincides with the prediction of the equilibrium SCGLE theory (and, qualitatively, with the MCT prediction). As already explained, this limiting scenario involves the divergence of $\eta^{*eq}(1/T)$ as the temperature $T$ approaches $T^a$ from above, and predicts the value $\eta^{*eq}(1/T)=\infty$ for temperatures $T<T^a$. This scenario is, of course, unobservable since it requires the impossible waiting time $t=\infty$. What is observed in any real experiment or simulation, instead, is a steep but finite increase of $\eta^*(t;\phi,T)$ with decreasing $T$, a behavior not yet explained satisfactorily. The results in Fig. \ref{fig:2} for finite $t$, however, provide a first but solid fundamental explanation.

Fig. \ref{fig:2} illustrates the theoretically-predicted scenario for finite waiting times. For any fixed $t$, the function $\eta^*(t;1/T_f)$ clearly exhibits two regimes as a function of $1/T$, separated by a  loosely-defined $t$-dependent crossover temperature $T_{co}(t)$. The first corresponds to samples that have fully equilibrated within the time $t$ (and hence, follow the plot of $\eta^{*eq}(1/T_f)$), and the second to samples that, at time $t$, have not yet reached equilibrium and continue aging. The red solid circle in Fig. \ref{fig:2}, for example, represents the crossover point $(1/T_{co},\eta^*(t;1/T_{co}))$ for the longest waiting time in the figure, $t=10^4$. Thus, the passage from the high-$T_f$ equilibrium regime to the low-$T_f$ non-equilibrium regime is clearly not an abrupt transition, but a soft crossover between them.  One notices that the crossover temperature $T_{co}(t)$ decreases with $t$, approaching asymptotically the value $T^a$ as $t\to\infty$.


The main message conveyed by Fig. \ref{fig:2}, thus, is that Eqs. \eqref{deltaeta}, along with \eqref{relsigmadif2appendix}-\eqref{lambdadk}, provide a route to the first-principles approximate  prediction of the evolution with waiting time $t$ of the $T$-dependence of $\eta^*(t;1/T)$ for realistic models of structural glass formers. As illustrated in the figure, these predictions at finite $t$ are completely free from unobservable divergences, exhibiting only a smooth crossover from the equilibration regime, at high-$T$, to a low-$T$ aging regime, in  which the system remains out of equilibrium  at time $t$. Thus, the kinetic perspective of the NE-SCGLE theory naturally reconciles the theoretical limiting MCT equilibrium scenario, which predicts an unobservable divergence, with the real experimental reports of the $T$-dependence of $\eta^*(t;1/T)$, often cast in the format of the \emph{Angell plots} of structural glass formers. Let us stress that this qualitative scenario was already  advanced since the early applications of the NE-SCGLE theory to the description of the aging of hard- and soft-sphere fluids, but the properties studied in those previous studies were the instantaneous mobility $b(t)$ and the $\alpha$-relaxation time $\tau_\alpha(t)$ \cite{nescgle3,mendoza,nescgle5}. Here we have complemented those previous studies with their rheological counterpart, represented by the behavior of the function $\eta^*(t;\phi,T)$ at a fixed volume fraction $\phi$.


\subsection{Time-dependent glass transition diagrams}\label{subsection4.2}

We have just seen how the first-principles NE-SCGLE-predicted behavior of an ensemble of systems quenched at many different final temperatures provides an enriched kinetic extension of the concept of \emph{Angell plots} \cite{angell}. One should then expect that the analysis of the behavior of the function $\eta^*(t;\phi,T)$, not at fixed packing fraction, but as a function of the waiting time $t$, the temperature $T$, and $\phi$, should provide an even richer kinetic scenario of the glass transition. In fact, such analysis may be a useful tool for the fundamental interpretation of the huge amount of experimental studies of glasses and of glass-forming processes, which frequently include explicitly the waiting time $t$ as an essential variable. For example, although cast in rather unfamiliar formats compared with the ordinary equilibrium phase diagrams, the so-called time-temperature-transformation (TTT) diagrams \cite{callisterrethwisch} are widely employed to describe the non-equilibrium transformation of a liquid after a cooling process, when it  passes through some of its equilibrium phases and non-equilibrium arrested states. Practical examples range from the experimental phase transformations in the manufacture of hard materials, such as iron and steel \cite{callisterrethwisch,atlasttd} and porous glasses \cite{nakashima}, but also include gel or glass formation by clays \cite{ruzicka1,ruzicka2} or proteins \cite{cardinaux,gibaud} in  solution. This relevant experimental phenomenology, thus, will surely benefit from the development of theoretical approaches that allow its first-principles understanding.

In this regard, let us mention that in a recent pioneering work \cite{zepeda} the predictions of the NE-SCGLE theory were precisely cast in the format of TTT diagrams, with the intention to build a first-principles description of these empirical diagrams. Ref.  \cite{zepeda}, however,  carried out such exercise in terms of the mobility function $b(t)$ defined in Eq. \eqref{bdt}, and for a different model system. Hence, in this section we start the application of our main contribution, Eq. \eqref{deltaeta}, to cast the resulting viscoelastic behavior of our simple WCA glass forming model in the format of a TTT diagram. For this, let us repeat the procedure of the  previous subsection and evaluate $\eta^*(t;\phi,T)$, but this time, not only for a single set of isochoric quenches along the isochore $\phi=0.7$, but for an enlarged ensemble of quenches, in which this restriction has been removed. More concretely, let us imagine that we prepare an ensemble of samples, all of them assumed originally in thermodynamic equilibrium (for times $t<0$), at different volume fractions $\phi$ but with the same initial temperature $T_i=10$. At $t=0$, all these equilibrium samples are instantaneously and isochorically quenched to an ensemble of finite final temperatures $T(<T_i)$, which are kept fixed afterwards ($t>0$). We then analyze the evolution of $\eta^*(t;\phi,T)$ for the whole ensemble, with the aim of describing the emerging time-dependent collective scenario.

\begin{figure}
	\includegraphics[width=0.32\textwidth, height=0.33\textwidth]{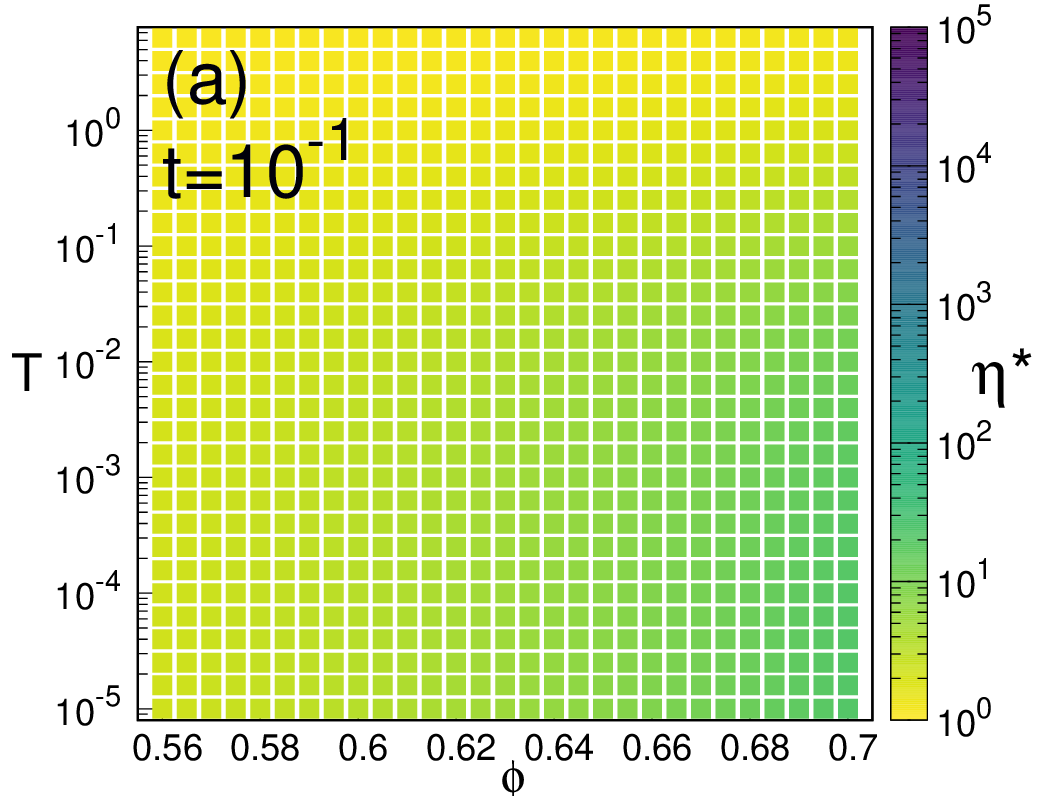}
	\includegraphics[width=0.32\textwidth, height=0.33\textwidth]{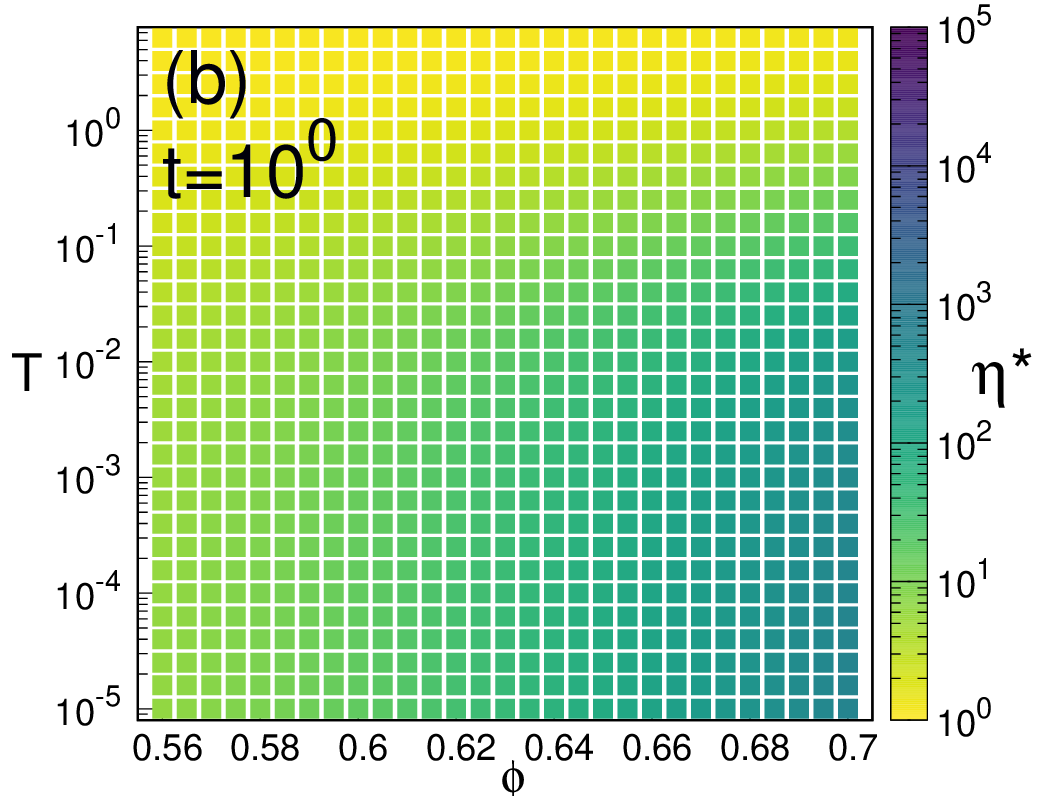}
	\includegraphics[width=0.32\textwidth, height=0.33\textwidth]{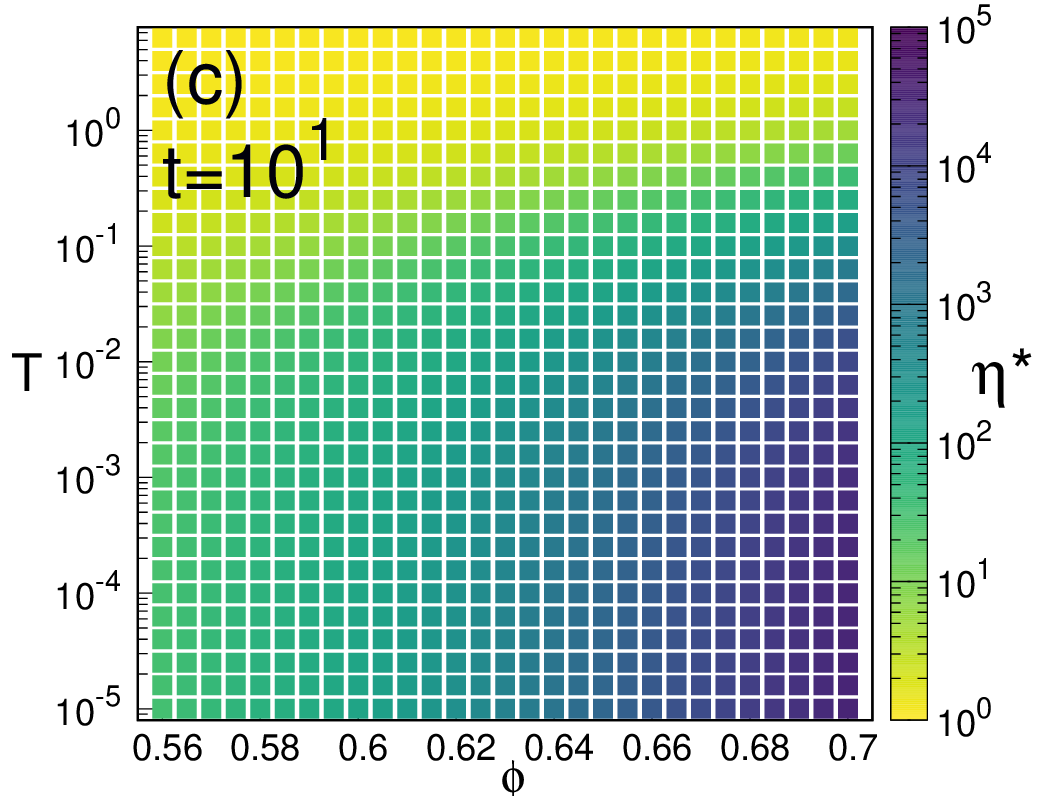}\\
	\includegraphics[width=0.32\textwidth, height=0.33\textwidth]{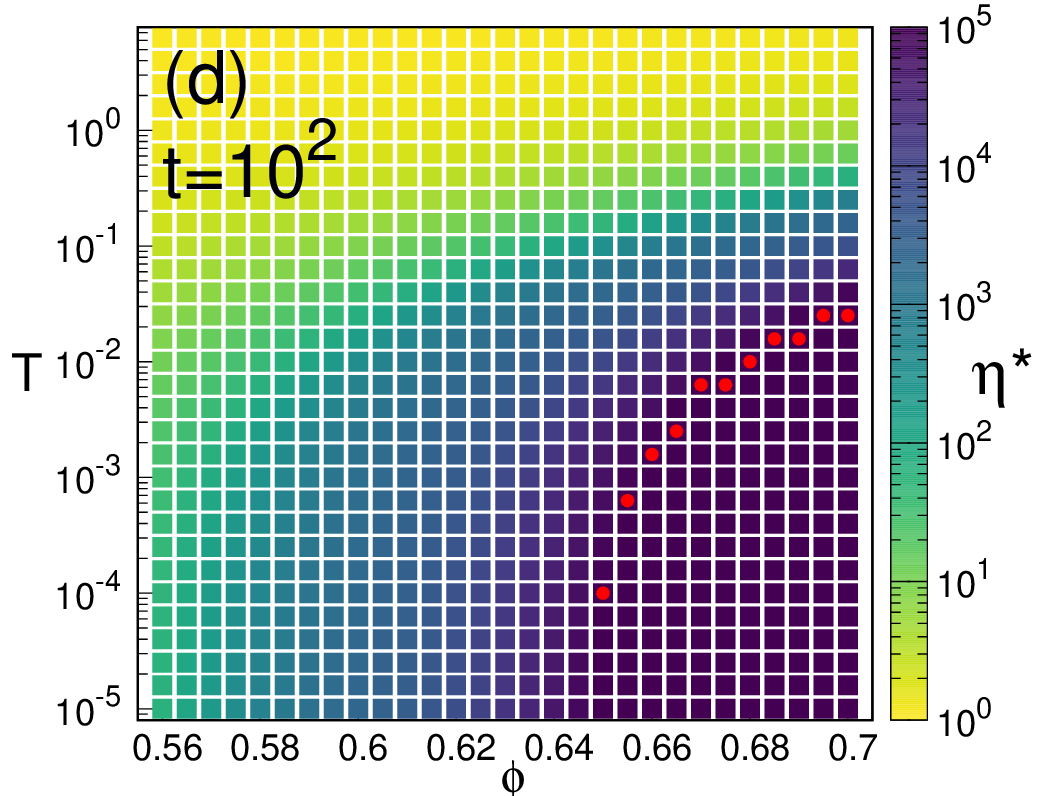}
	\includegraphics[width=0.32\textwidth, height=0.33\textwidth]{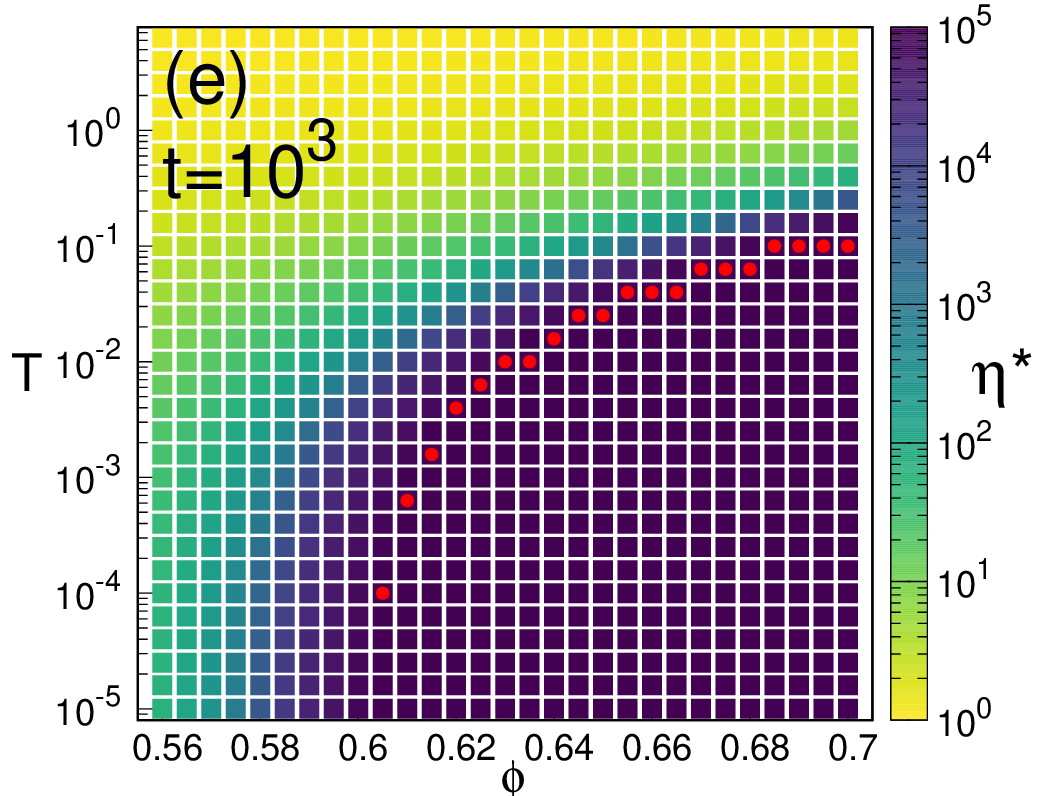}
	\includegraphics[width=0.32\textwidth, height=0.33\textwidth]{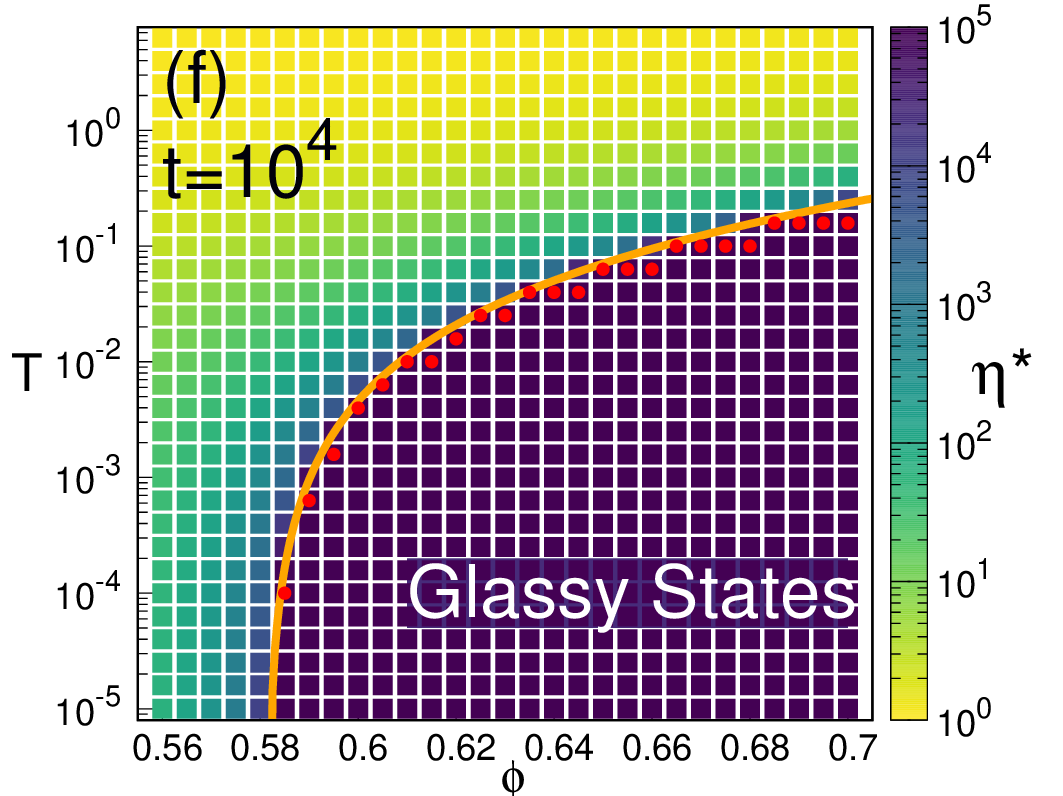}
	\caption{Sequence of snapshots describing the time evolution of the instantaneous viscosity $\eta^*(\phi,T;t)$ along the $(\phi,T_f)$-plane, for a suddenly quenched WCA liquid at $t=10^{-1}$ (a), $t=10^{0}$ (b), $t=10^{1}$ (c), $t=10^{2}$ (d), $t=10^{3}$ (e), $t=10^{4}$ (f). In all the cases, the instantaneous value of $\eta^*(\phi,T;t)$ is represented by the color code depicted on the right side of each figure. The solid line in (f) describes the locus of the boundary $\eta^*[\phi;T]=10^5$ which separates the ergodic region from the domain of glassy states (see the text)}. 
	\label{fig:3}
\end{figure}

The six panels of Fig. \ref{fig:3} summarize the results obtained for the set of final state points indicated there. Each of these panels corresponds to a snapshot of the ($\phi;T$)-plane at different waiting times, in which the instantaneous value of the viscosity is represented, at each state point, using the chromatic code on the right side. Specifically, the snapshots correspond to $t=10^{-1}, 10^0, 10^1, 10^2, 10^3$ and $10^4$, which thus depict the intermediate- to long-time stages of all the quench processes considered. The first feature to observe in Fig. \ref{fig:3} is the overall monotonic increase in $\eta^*(t;\phi,T)$ for waiting times $10^{-1}<t<10^{1}$ (upper panels), occurring faster in the region of large volume fractions and low temperatures. This is visualized through the progressive darkening of the state points, starting from the bottom-right region of the $(\phi,T)$-plane and gradually spreading to lower densities and higher temperatures. 

At longer times (say, $t\ge 10^2$) the viscosity $\eta^*(t;\phi,T)$ increases with very heterogeneous rates, depending strongly on the final state point  $(\phi,T)$, and outlining essentially two well-defined regions. In the low-density high-temperature (yellow and green) equilibrium region, $\eta^*(t;\phi,T)$ has reached, or soon reaches, its constant and finite  equilibrium value $\eta^{*eq}(\phi,T)$. In the opposite corner (high densities and low temperatures) $\eta^*(t;\phi,T)$ continues to increase without bound, eventually growing above any arbitrary threshold $\eta^*_g$. Recall, for example, the value $\eta^*_g=10^{16}$ defining the  empirical glass transition temperature of molecular liquids. In analogy, the present theory allows us to draw a $t$-dependent empirical  glass transition line $T=T_g(\phi;t)$, defined by the glass-transition condition  $\eta^*(t;\phi,T_g)=\eta^*_g$. This empirical glass transition line, represented in Figs. \ref{fig:3}(d)-\ref{fig:3}(f) by the red dots, approaches its long-$t$ stationary limit  $T=T_g(\phi;t\to\infty)$ at an increasingly slower rate (indicated, for clarity, by the solid line in Fig. \ref{fig:3}(f)). These illustrative results were calculated using a glass-transition threshold value of $\eta^*_g=10^{5}$  (rather than  $\eta^*_g=10^{16}$), more adequate for the characteristic values of a colloidal suspension of repulsive particles approaching the GT \cite{cheng,weeks}. In this manner, the $t$-dependent glass transition line $T=T_g(\phi;t)$ separates the state space $(\phi,T)$ at time $t$ in two mutually exclusive regions: one in which the value of the viscosity remains smaller than the threshold $\eta^*_g=10^{5}$, so that the WCA system remains in a liquid state, and  another in which $\eta^*(t;\phi,T)$ is  larger than $10^5$,  identified with the glass region.

Let us finally mention that, notwithstanding the fact that the above results refer to the particular case of the WCA model, they provide useful insights on the universal nature of the experimental glass transition observed in colloidal suspensions governed by short-ranged repulsive interactions, such as those acting between hard-spheres or PNIPAM particles \cite{rivas,cheng,weeks}, in which empirical criteria involving a finite threshold value for the viscosity are commonly used. As already mentioned, this empirical definition contrasts with the theoretical prediction of the divergence of the viscosity (or, equivalently, of the structural relaxation time $\tau^{\alpha}$) when crossing the \emph{ideal} glass transition (or dynamic arrest) line $T=T^a(\phi)$, defined by the condition $\eta^{*eq}(\phi,T^a)=\infty$, first predicted by MCT and by the equilibrium SCGLE theory. In this regard, one of the farthest-reaching contributions of this work (and of the NE-SCGLE theory, in general) is the clarification of the relationship between this \emph{ideal} glass transition line $T=T^a(\phi)$ and the $t$-dependent \emph{empirical}  glass transition line $T=T_g(\phi;t)$, defined by the  condition  $\eta^*(t;\phi,T_g)=\eta^*_g$: the latter becomes the former when $\eta^*_g=\infty$ and  $t=\infty$.

\section{Summary and concluding remarks}\label{section5}

In this work we have proposed a general theoretical framework for the first-principles prediction of the non-equilibrium linear viscoelastic response of a glass- or gel-forming colloidal  liquid that has been suddenly quenched from arbitrary ergodic initial conditions, to an arbitrary final density and temperature. Given that the quench occurred at time $t=0$, this non-equilibrium response refers to the evolution of the viscoelastic properties at later waiting times $t>0$. The referred theoretical protocol results from merging the fundamental equations of the  NE-SCGLE theory (whose solution yields the time-evolution of the non-equilibrium structure factor $S(k;t)$ and collective and self  ISFs, $F(k,\tau;t)$ and $F_S(k,\tau;t)$), with the general but approximate expression for the total shear-stress relaxation function $\eta(\tau;t)$, Eq. (\ref{deltaeta}),  in terms of  $S(k;t)$ and $F(k,\tau;t)$.     

The most relevant contribution of this work is the derivation of Eq. \eqref{deltaeta}. For this we adopted the simplest statistical mechanical strategy, namely, to check that the arguments, simplifications, and approximations employed by  N\"agele and Bergenholtz \cite{naegele,banchio} for equilibrium Brownian liquids never required the condition of thermodynamic equilibrium, but only mathematical symmetry conditions, such as  stationarity or other temporal or spatial symmetries. Leaving some details for a separate and related publication that focused precisely on non-equilibrium statistical mechanical methodologies \cite{wertheimlovett}, here we could concentrate in illustrating the concrete use of the general theoretical protocol presented here for the prediction of the aging of the linear viscoelasticity in non-equilibrium glass-forming liquids.

Thus, the second most relevant contribution of this work is the systematic application of Eq. \ref{deltaeta} with Eqs. \eqref{relsigmadif2appendix}-\eqref{lambdadk}, to the description of the rheological behavior of the WCA soft-sphere model fluid. We first illustrated the  non-equilibrium evolution of the relaxation function $\Delta\eta(\tau;t)$ for individual quench processes, comparing the process of equilibration with the aging process of dynamic arrest. We then described the scenario revealed by considering the collective evolution of an ensemble of quench processes with many final densities and temperatures. We first considered an ensemble of quenches with fixed volume fraction, to discuss the enriched concept of $t$-dependent Angell plots, and then extended this exercise to the case in which the volume fraction is no longer fixed, thus leading to the recently-developed concept of time-dependent glass-transition diagrams \cite{zepeda}, applied here in the context of the non-equilibrium linear viscoelasticity of the WCA model.

Among the most salient features of the present work, let us first mention some general theoretical considerations. According to Eq. \eqref{deltaeta}, $S(k;t)$ and $F(k,\tau;t)$ are the main microscopic elements that determine the instantaneous value of the viscosity. The NE-SCGLE equations \eqref{relsigmadif2appendix}-\eqref{lambdadk}, in their turn, allow us to directly relate the viscoelastic response of the system with explicit microscopic details, such as the potential of interaction between the constituent particles, and the protocol of preparation. We refer to the arbitrary ergodic initial state and the arbitrary final density and temperature, in the simplified example involving an instantaneous quench. However, considering other preparation protocols, involving, for example, an arbitrary temperature program $T=T(t)$, could be approximated by a sequence of differential instantaneous quenches.  

The resulting theoretical approach is now ready for a series of systematic applications to characterize the viscoelastic response in a diversity of qualitatively different glass- and gel-forming  systems characterized, for example, by Lennard Jones-like interactions or systems with competing interactions (short-ranged attraction plus long-ranged repulsion), in which the competence between thermodynamical instabilities (spinodal line, $\lambda$-line) and dynamical arrest mechanisms leads to the possibility of qualitatively different glassy states, ranging from porous glasses, gels and Wigner glasses \cite{carretas}. 

There are also no fundamental barriers that prevent the extension of the arguments and equations presented here to much more complex conditions, involving glass and gel forming systems with multiple relaxation channels. This is the case, for example, of colloidal suspensions comprised by dipolar particles (ferrofluids), in which the decoupling of the orientational and translational dynamics allows one to investigate partially arrested states, and also, of mixtures with disparate size ratios, which allow for the development of qualitatively different glassy states upon tuning the molar distribution. 

Finally, a relevant contribution of the present and previous work is to insist on the relevant fundamental  connection between the experimental glass transition and the conventional MCT theoretical description, severely restricted to the asymptotic limit $t\to\infty$. Therefore, the results presented here, extend those equilibrium predictions regarding the existence of a critical temperature $T ^g(\phi)$ predicting the divergence of $\eta(T)$,  to a more realistic scenario, predicted for any finite time $t$ after the quench, thus providing a richer and more realistic conceptual scenario, that can be employed to understand the phenomenology of aging and glassy dyamics in real and simulated materials, which only occur within the finite time-span of any practical observation. This, in turn, allows us to provide meaningful theoretical predictions regarding the viscoelastic response of gelled and glassy materials  at any finite waiting times, starting from the experimentally accessible short and intermediate times, but extending to the often inaccessible asymptotic long times.

\medskip
\section{Acknowledgments}
ACKNOWLEDGMENTS: This work was supported by the Consejo Nacional de Humanidades, Ciencias y Tecnolog\'ias (CONAHCYT, Mexico) through Postdoctoral Fellowships Grants No. I1200/224/2021 and I1200/320/2022; and trough grants 320983, CB A1-S-22362. This work was also supported by the Laboratorio Nacional de Ingenier\'ia de la Materia Fuera de Equilibrio (LANIMFE).

\vskip5cm



\begin{thebibliography}{99}

\bibitem{mewiswagnerbook21} Theory and Applications of Colloidal Suspension Rheology, edited by N. J. Wagner and J. Mewis (Cambridge University Press, Cambridge, UK, 2021).

\bibitem{ChenJanmeyYodh} D.T.N. Chen, Q. Wen, P. Janmey, J. Crocker, ans A. Yodh, Ann. Rev. Condens. Matter Phys. 1, 301 (2010).

\bibitem{johari} G.P. Johari, \emph{Thermochimica Acta} \textbf{693} (2020) 178715.


\bibitem{voigtmann1} Th. Voigtmann, \emph{Curr. Opin. Colloid Interface Sci.} Volume 19, Issue 6, 549-560 (2014).

\bibitem{winkler} R.G. Winkler, D.A. Fedosov, G. Gompper, \emph{Curr. Opin. Colloid Interface Sci.} Volume 19, Issue 6, 594-610 (2014). 


\bibitem{ferry} J.D. Ferry, {\it Viscoelastic Properties of Polymers}, Wiley, New York (1980).

\bibitem{purnomo} E.H. Purnomo, D. van den Ende, J. Mellema and F. Mugele, \emph{Europhys. Lett.}, \textbf{76} (1), pp. 74-80 (2006).

\bibitem{purnomo1} E.H. Purnomo, D. van den Ende, S.A. Vanapalli and F. Mugele \emph{Phys. Rev. Lett} \textbf{101} 238301 (2008).

\bibitem{christopoulou} C. Christopoulou, G. Petekidis, B. Erwin, M. Cloitre, And D. Vlassopoulos, Phil. Trans. R. Soc. A  \textbf{367}, 5051 (2009).

\bibitem{suman} K. Suman and N. G. Wagner, J. Chem. Phys. \textbf{157}, 024901 (2022).

\bibitem{mckenna1} J. Zhao, S.L. Simon and G.B. McKenna, \emph{Nat. Commun.} \textbf{4}, 1783 (2013).

\bibitem{mckenna2} X. Shi, A. Mandanici and G.B. McKenna, \emph{J. Chem. Phys.} \textbf{123} 174507 (2005).

\bibitem{mckenna3} Polymer Engineering \& Science Vol. \textbf{62}, Issue 5, 1321-1730 (2022).

\bibitem{geszti} T. Geszti, J. Phys. C 16, 5805 (1983). 

\bibitem{naegele} G. N\"agele and J. Bergenholtz, \emph{J. Chem. Phys.} \textbf{108} 9893 (1998).

\bibitem{banchio} A.J. Banchio, G.N\"agele and J. Bergenholtz \emph{J. Chem. Phys.} \textbf{111}, 8721 (1999).

\bibitem{goetze1} W. G\"otze and L. Sj\"ogren, \emph{Rep. Prog. Phys.} \textbf{55}, 241 (1992).

\bibitem{brader} J.M. Brader, Th. Voigtmann, M. Fuchs and M.E. Cates, \emph{Proc. Natl. Acad. Sci.} \textbf{106} (36) 15186-15191.

\bibitem{brady} J.Brady \emph{J. Chem. Phys.} \textbf{99}, 567-581 (1993).

\bibitem{mcquarrie}  D. A. McQuarrie {\em Statistical Mechanics}, Harper \& Row (New York, 1976).

\bibitem{hansen}  J.-P. Hansen and I. R. McDonald, {\em Theory of Simple 
Liquids} (Elsevier-Academic Press, 2006).

\bibitem{nescgle1} P. E. Ram\'irez-Gonz\'alez and M. Medina-Noyola, \emph{Phys. Rev. E} \textbf{82}, 061503 (2010).

\bibitem{nescgle2} P. E. Ram\'irez-Gonz\'alez and M. Medina-Noyola \emph{Phys. Rev. E} \textbf{82}, 061504 (2010).

\bibitem{nescgle3} L.E. S\'anchez-D\'iaz, P. Ram\'irez-Gonz\'alez, and M. Medina-Noyola
\emph{Phys. Rev. E} \textbf{87}, 052306 (2013). 



\bibitem{gabriel} G. Perez, et al. \emph{Phys. Rev. E} \textbf{83}, 060501(R) (2011).

\bibitem{nescgle5} P. Mendoza-M\'endez,E. L\'azaro-L\'azaro,
L. E. S\'anchez-D\'iaz, , P. E. Ram\'irez-Gonz\'alez,
G. P\'erez-\'Angel,  and M. Medina-Noyola, Physical Review E {\bf 96},
022608 (2017).

\bibitem{mendoza} P. Mendoza-M\'endez et.al. \emph{J. Chem. Phys.} \textbf{157}, 244504 (2022). 

\bibitem{evans} R. Evans \textit{Adv. Phys.} \textbf{28} 143-200 (1979).

\bibitem{wertheimlovett} O. Joaqu\'in-Jaime,  R. Peredo-Ortiz,  M. Medina-Noyola, and L.F. Elizondo-Aguilera, ``From equilibrium to non-equilibrium statistical mechanics of liquids'', manuscript submitted to \emph{Phys. Rev. E} (2024) / https://doi.org/10.48550/arXiv.2401.15220.

\bibitem{wca} H.C. Andersen, J.D. Weeks and D. Chandler, \emph{Phys. Rev. A} \textbf{4}, 1597 (1971).

\bibitem{angell} C.A. Angell \emph{Science} \textbf{267}: 1924 (1995).

\bibitem{zepeda} J.B. Zepeda-LÃ³pez and M. Medina-Noyola, \emph{J. Chem. Phys.} \textbf{154}, 174901 (2021).

\bibitem{prlhi} M. Medina-Noyola, Phys. Rev. Lett. {\bf 60}, 2705 (1988).

\bibitem{mazurgeigen} P. Mazur and U. Geigenm\"uller, Physica A \textbf{146},
657 (1987).

\bibitem{boonyip}  J. L. Boon and S. Yip, {\em Molecular Hydrodynamics}{\it \ }%
(Dover Publications Inc. N. Y., 1980).

\bibitem{nonlinonsmach} Ricardo Peredo-Ortiz, Luis F. Elizondo-Aguilera, Pedro Ram\'irez- Gonz\'alez, Edilio L\'azaro-L\'azaro, Patricia Mendoza-M\'endez and Magdaleno Medina-Noyola (26 Dec 2023): \emph{Non-equilibrium Onsager-Machlup theory}, Molecular Physics, DOI: 10.1080/00268976.2023.2297991.

\bibitem{rivas} R. Rivas-Barbosa, E. L\'azaro-L\'azaro, P. Mendoza-Mendez, Tim Still,
V. Piazza, P.E. Ram\'irez-Gonz\'alez, M. Medina-Noyola and M. Laurati.

\bibitem{percus}J. K. Percus and G. J. Yevick, Phys. Rev. \textbf{110}, 1 (1957).

\bibitem{verlet} L. Verlet and J.-J. Weis, \emph{Phys. Rev. A}
\textbf{5}, 939 (1972).

\bibitem{elizondo_prl} L.F. Elizondo-Aguilera, T. Rizzo, and Th. Voigtmann \emph{Phys. Rev. Lett.} \textbf{129} 238003 (2022).

\bibitem{06Likos} C. N. Likos, Soft Matter \textbf{2}, 478 (2006).


\bibitem{dyneq0}F. de J. Guevara-Rodr\'iguez and M.Medina-Noyola,Phys. Rev. E {68}, 011405 (2003).

\bibitem{dyneqPRL} P. E. Ram\'irez-Gonz\'alez, L. L\'opez-Flores, H. Acu\~na-Campa, and M. Medina-Noyola,  Phys. Rev. Lett. \textbf{107}, 155701 (2011).

\bibitem{dyneqPRE1} L. L\'opez-Flores, H. Ru\'iz-Estrada, M. Ch\'avez-P\'aez,  and M. Medina-Noyola, Phys. Rev. E. {88}, 042301 (2013).

\bibitem{dyneqPRE2} L. L\'opez-Flores, J. M. Olais-Govea, M. Ch\'avez-P\'aez,  and M. Medina-Noyola, Phys. Rev. E. \textbf{103}, L050602 (2021).

\bibitem{callisterrethwisch} W. D. Callister \& D. G. Rethwisch. {\em \ Materials Science and Engineering: An Introduction}. 10th edition, John Wiley  (2018).

\bibitem{ruzicka1} B. Ruzicka  et al. Nature Materials \textbf{10}, 56 (2011).

\bibitem{huangmckenna} D. Huang and G. B. McKenna, J. Chem. Phys. 114, 5621 (2001).

\bibitem{kelton} K. F. Kelton, J. Phys.: Condens. Matter 29, 023002 (2017).

\bibitem{mauroallan} J. C. Mauro, D. C. Allan, and M. Potuzak, Phys. Rev. B 80, 094204 (2009).

\bibitem{berthierbiroli} L. Berthier and G. Biroli, Rev. Mod. Phys. 83, 587 (2011).

\bibitem{tarjus} G. Tarjus, ``An overview of the theories of the glass transition'' in \emph{Dynamical Heterogeneities in Glasses, Colloids, and Granular Media}, L. Berthier, G. Biroli, J.-P. Bouchaud, L. Cipelletti, W. van Saarloos, Eds. (Oxford University Press, New York, NY, 2011), pp.39?67.

\bibitem{jensen} S. Ciarella, R. A. Biezemans, and L. M. C. Janssen, ``Understanding, predicting, and tuning the fragility of vitrimeric polymers''. Proc Natl Acad Sci U S A. 116(50):25013-25022 (2019). 

\bibitem{todos1} L. Yeomans-Reyna, M. A. Ch\'avez-Rojo, P. E. Ram\'{\i}rez-Gonz\'alez, R. Ju\'arez-Maldonado, M. Ch\'avez-P\'aez, and M. Medina-Noyola, Phys. Rev. E {\bf 76}, 041504 (2007)

\bibitem{reviewroque} R. Ju\'arez-Maldonado and M. Medina-Noyola, in {\it Structure and functional properties of colloidal systems}, Ed. R. Hidalgo, Surfactant Science Series, Vol. 104. (ISBN 978-1-4200-8446-7, CRC Press Taylor \& Francis, 2009). 

\bibitem{eliz_voigt} L.F. Elizondo-Aguilera and Th. Voigtmann, \emph{Phys. Rev. E} \textbf{100}, 042601 (2019).


\bibitem{atlasttd} G. F. Vander Voort. {\em \ Atlas of Time Temperature Diagrams for Irons and Steels.} ASM INTERNATIONAL, Materials Park, OH (1991).

\bibitem{nakashima} K. Nakashima, K. Noda, and K. Mori, J. Am. Ceram. Soc., \textbf{80}, 1101 (1997).

\bibitem{ruzicka2} B. Ruzicka \& E. Zaccarelli. Soft Matter \textbf{7}, 4800--4805 (2011).

\bibitem{cardinaux} F. Cardinaux, T. Gibaud, A. Stradner, and P. Schurtenberger, Phys. Rev. Lett. \textbf{99}, 118301 (2007).

\bibitem{gibaud}   T. Gibaud and P. Schurtenberger, J. Phys.:
Condens. Matter \textbf{21}, 322201 (2009).

\bibitem{cheng} Z. Cheng, J. Zhu, P.M. Chaikin, S-E- Phan, and W.B. Russel
Phys. Rev. E \textbf{65}, 041405 (2002).

\bibitem{weeks} G.L. Hunter and E.R. Weeks, \emph{Rep. Prog. Phys.} \textbf{75}, 066501 (2012).

\bibitem{carretas} A.G. Carretas-Talamante, L.F. Elizondo-Aguilera, J.B. Zepeda-L\'opez, E. L\'azaro-L\'azaro and M. Medina-Noyola, \emph{J. Chem. Phys.} \textbf{158}, 064506 (2023).


\end{thebibliography}
\end{document}